\documentclass[twocolumn,reprint,amsmath,amssymb,aps,superscriptaddress]{revtex4-2}

\usepackage{amsmath}
\usepackage{amssymb}
\usepackage{MnSymbol}
\usepackage{graphicx}
\usepackage[colorlinks=true, allcolors=blue]{hyperref}
\usepackage{siunitx}
\usepackage{float} 
\usepackage{natbib}
\usepackage{MnSymbol}
\usepackage{hyperref}
\usepackage{bm}
\usepackage{dcolumn}
\usepackage{tabstackengine}
\usepackage[dvipsnames]{xcolor}
\usepackage{ulem}
\usepackage{upgreek}
\usepackage{mathtools}
\usepackage{tabularx}

\newcolumntype{L}[1]{>{\hsize=#1\hsize\raggedright\arraybackslash}X}%
\newcolumntype{R}[1]{>{\hsize=#1\hsize\raggedleft\arraybackslash}X}%
\newcolumntype{C}[1]{>{\hsize=#1\hsize\centering\arraybackslash}X}%

\makeatletter
\newcommand*{\balancecolsandclearpage}{%
  \close@column@grid
  \clearpage
  \twocolumngrid
}
\makeatother

\stackMath{}

\newcommand{\dd}[2]{\frac{\mathrm{d}#1}{\mathrm{d} #2}}
\newcommand{\pd}[2]{\frac{\partial #1}{\partial #2}}

\newcommand{\sub}[1]{_{\textrm{#1}}}

\newcommand{\vect}{\boldsymbol}
\newcommand{\Eqref}[1]{\mbox{equation\hspace{0.25em}\eqref{#1}}}
\newcommand{\Eqsref}[1]{\mbox{equations\hspace{0.25em}\eqref{#1}}}

\newcommand{\figref}[1]{\mbox{figure\hspace{0.25em}\ref{#1}}}
\newcommand{\Figref}[1]{\mbox{Figure\hspace{0.25em}\ref{#1}}}
\newcommand{\figsref}[1]{\mbox{figures\hspace{0.25em}\ref{#1}}}

\newcommand{\tabref}[1]{\mbox{table \hspace{0.25em}\ref{#1}}}

\newcommand{\Vc}{V_\mathrm{tot}}

\newcommand{\rc}{r_\textrm{c}}
\newcommand{\Vf}{V_{\mathrm{f}}}
\newcommand{\Vcore}{V_{\mathrm{c}}}
\newcommand{\KB}{K_\mathrm{b}}
\newcommand{\KS}{K_\mathrm{s}}
\newcommand{\VCrit}{\epsilon \sub{crit}}
\newcommand{\Supp}{Supplementary Information}
\newcommand*\diff{\mathop{}\!\mathrm{d}}

\begin{document}

\title{Active viscoelastic condensates provide controllable mechanical anchor points}

\author{Oliver W. Paulin}
\affiliation{Max Planck Institute for Dynamics and Self-Organization, Am Fa{\ss}berg 17, 37077 G\"{o}ttingen, Germany}
\author{Júlia Garcia-Baucells}
\affiliation{Max Perutz Labs, University of Vienna, Vienna Biocenter (VBC); Vienna, A-1030, Austria}
\affiliation{Vienna BioCenter PhD Program, Doctoral School of the University of Vienna and Medical University of Vienna; Vienna, A-1030, Austria}
\author{Luise Zieger}
\affiliation{Institute of Numerical Mathematics and Optimization, TU Bergakademie Freiberg, Akademiestr. 6, 09599 Freiberg, Germany}
\affiliation{Faculty of Computer Science/Mathematics, HTW Dresden, Friedrich-List-Platz 1, 01069 Dresden, Germany}
\author{Sebastian Aland}
\affiliation{Institute of Numerical Mathematics and Optimization, TU Bergakademie Freiberg, Akademiestr. 6, 09599 Freiberg, Germany}
\affiliation{Faculty of Computer Science/Mathematics, HTW Dresden, Friedrich-List-Platz 1, 01069 Dresden, Germany}
\affiliation{Center for Systems Biology Dresden, Pfotenhauerstr. 108, 01307 Dresden, Germany}
\author{Alexander Dammermann}
\affiliation{Max Perutz Labs, University of Vienna, Vienna Biocenter (VBC); Vienna, A-1030, Austria}
\author{David Zwicker}
\affiliation{Max Planck Institute for Dynamics and Self-Organization, Am Fa{\ss}berg 17, 37077 G\"{o}ttingen, Germany}

\date{\today}

\begin{abstract}
Many biological materials must couple mechanical strength with the ability to rapidly self-assemble at a specific location. 
In particular, biomolecular condensates readily self-assemble via phase separation, but may also need to resist external forces to fulfil their function.
Spatial localisation of condensate formation can be controlled by active cores that preferentially drive the production of condensate material at a particular point, while resistance to external forces can be facilitated by viscoelastic material properties. 
To investigate the interplay of these two processes, we develop a continuum model of viscoelastic growth around an active core. 
We find that viscoelastic stresses restrict condensate growth, but also impart resistance to deformation.
We investigate the effect of different incorporation schemes on growth dynamics, and test the influence of mechanical properties on condensate strength.
Finally, we compare the predictions of our model to experimental data from centrosomes in \textit{C. elegans} embryos, identifying a parameter regime in which rapid growth can be combined with appropriate mechanical strength, and studying how strain-dependent material incorporation may lead to isotropic growth of scaffold material.
Our results provide general design principles for other materials that must reconcile rapid, localised self-assembly with mechanical strength, such as focal adhesions.
\end{abstract}

\maketitle

\section{Introduction}

\begin{figure*} 
    \centering
    \includegraphics[width=\textwidth]{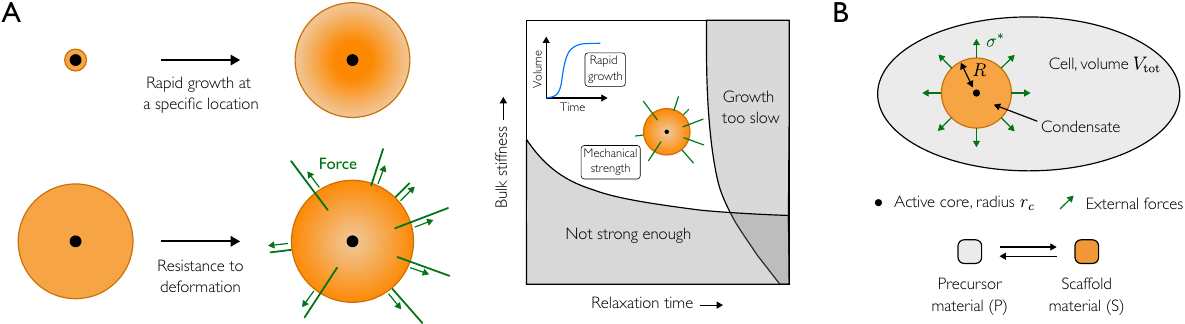}
    \caption{(A) Various biological and synthetic materials must exhibit both rapid formation at a specific location and mechanical resistance to external deformations. A viscoelastic material model coupled with incorporation via active chemical reactions enables both of these behaviours in an appropriate parameter regime.
    (B) Schematic highlighting our modelling framework, in which a viscoelastic condensate of radius $R$ forms around an active core of radius $\rc$. The condensate is embedded within a cell of volume $\Vc$, and its growth is controlled by a conversion reaction between a precursor component $P$, and its corresponding assembly-competent (or scaffold) form $S$. The effect of external forces on condensate growth is quantified by a radial stress~$\sigma^*$.
    }
    \label{fig:Overview}
\end{figure*}

Biological cells continuously form transient structures at specific locations in space and time to fulfil specific functions.
Biomolecular condensates are a widespread example of such structures, and are involved in almost all sub-cellular processes, including sequestering molecules~\cite{Bracha2019}, catalysing reactions~\cite{Oflynn2021}, and mediating mechanical forces~\cite{Sanfeliu2025}.
Rapid condensate assembly is often achieved via phase separation resulting from weak multivalent interactions between constituent biomolecules~\cite{Brangwynne2009,Banani2017,Dignon2020}.
Alternatively, condensates may assemble as a lattice-like gel structure, mediated by specific interactions between molecules \cite{Yoo2019,Korkmazhan2021,Raff2019}.
For condensates whose function involves the anchoring of sustained external forces, such as centrosomes \cite{Raff2019,Woodruff2021}, focal adhesions \cite{Litschel2024}, and tight junctions \cite{Schwayer2019,Beutel2019,Sun2024}, the requirement of mechanical strength induces an additional design challenge (\Figref{fig:Overview}A): how can an object assemble rapidly in a liquid-like fashion, but also provide strong resistance to deformation, like an elastic solid? Further, how can this process be controlled to ensure that such objects only form exactly when and where they are supposed to?

Passive condensate formation, either by phase separation or lattice-like growth, is difficult to control; condensates typically form spontaneously throughout the cell, and may coarsen over time.
Conversely, condensate formation that is driven by active (\textit{i.e.}, energy-consuming) processes allows exquisite control over condensate size, positioning, timing, and number~\cite{Weber2019,Zwicker2025}.
In particular, spatial control can be enabled by catalytically active cores that drive the localised conversion of biomolecules from a soluble precursor form to a corresponding assembly-competent form, facilitating controlled nucleation~\cite{Zwicker2014}.
Condensate growth may then be accelerated by further conversion reactions within the condensate interior (\textit{e.g.}, due to preferential partitioning of enzymes into the condensate).
Internal incorporation of new material requires that condensates have a certain degree of fluidity. %
A pure liquid, however, will not provide the requisite mechanical strength in most applications.
Viscoelastic material properties, on the other hand, permit both solid-like mechanical resistance and liquid-like fluidity.
We therefore propose the combination of viscoelasticity and active material incorporation as key ingredients for the controllable formation of mechanical anchors in biological systems. %

In this manuscript, we construct a physical model of viscoelastic condensate growth via active incorporation. Our model is agnostic as to whether the condensate microstructure constitutes a gel-like lattice or phase-separated droplet, and enables predictions for how condensate size, density, and stress state evolve with time.
To highlight the utility of our model, we compare our theoretical predictions to experimental data for centrosome growth in \textit{C. elegans} embryos, constraining the material parameters which are consistent with both rapid centrosome maturation in the approach to mitosis, and the ability to anchor microtubule-mediated mitotic forces.

\section{Viscoelastic model of condensate growth}\label{sec:Model}

To construct our model, we assume that condensate material can exist in two distinct forms: a precursor form, and an assembly-competent (or scaffold) form.
Condensate material can be converted between these two configurations by active chemical reactions.
To ensure controllable condensate formation at specific locations in space, active cores drive the localised conversion of precursor material into scaffold material, which then rapidly self-assembles into a compact, condensed state.
Further condensate growth results from the incorporation of new material via additional conversion reactions in the condensate interior.
We describe the boundary between the condensate interior and the surrounding environment as a sharp interface across which precursor material can be exchanged.
For a porous gel-like material, this interface is defined as the edge of the connected scaffold network.
For a droplet-like phase-separated material, the interface emerges naturally from phase separation. %

To describe the distribution of precursor~($P$) and scaffold~($S$) components within a condensate, we take a continuum approach in which the relative volume fractions of $P$ and $S$ are given by $\phi_{P}(\vect{r},t)$ and $\phi_{S}(\vect{r},t)$, respectively.
Below, we derive a model that describes how $\phi_{P}$ and $\phi_{S}$ evolve within a condensate over time. %
\Figref{fig:Overview}B demonstrates an outline of our modelling framework.

\subsection{{The cell provides a finite reservoir of condensate material}} \label{sec:Precursor}

To begin, we assume that condensates are embedded within a closed system, such as a cell, of volume~$\Vc$.
The total amount of condensate material (precursor plus scaffold) available is fixed,  
${\int \left( \phi_{S}+\phi_{P} \right)\textrm{d}V = \bar{\phi}\Vc}$, where $\bar{\phi}$ is the mean volume fraction of condensate material in the cell.
For simplicity, we also assume that individual condensates grow independently from each other, and are only coupled via the shared pool of precursor material.
As such, we focus on the growth of a single spherical condensate of radius~$R$.

\subsection{Internal reactions fuel condensate growth} \label{sec:GrowthModel}

The evolution of scaffold material within the condensate is governed by the continuity equation
\begin{equation}\label{eq:ContMassBalance}
        \pd{ \phi_{S}}{t}+\nabla\cdot\vect{j}=s \; ,
\end{equation}
where $\vect{j}=\phi_S\vect{v}$ is the flux of scaffold material with velocity $\vect{v}$ relative to the centre-of-mass (lab) frame, and $s$ is a source term that captures the conversion of precursor components into scaffold material. 
The interface between the condensate and the rest of the cell is determined by the kinematic condition $\partial_t R =  \left[\vect{v} \cdot \hat{\vect{n}}\right]_{r=R}$ in a spherically symmetric system, where $\hat{\vect{n}}$ is the normal to the surface.
Here, we assume that the flux of scaffold material across this interface is negligible since we expect condensate growth to be dominated by \textit{internal} incorporation of new scaffold material. 

Incorporation by conversion reactions takes place either at the surface of the active core or within the bulk of the condensate, which we capture by taking
\begin{align}\label{eq:SourceGeneral}
    s&=\left[k_{+} +q_\mathrm{c} (\vect r)\right] {\phi}_{P} -k_{-} \phi_{S} \; .
\end{align}
Here, $k_+$ is the rate of conversion of precursor material into scaffold material in the bulk of the condensate; $k_-$ is the rate of turnover of scaffold back into precursor; and ${q _\mathrm{c}(\vect r)= q\delta(r - r_\mathrm{c})}$ quantifies the conversion of precursor into scaffold at the surface of the active core, which we parameterise as a rigid sphere with fixed radius $r_\mathrm{c}$. 
In general, $k_{+}$, $k_{-}$, and $q$ need not be constant, but may depend on the strain in the scaffold network.

\subsection{Material growth strains the scaffold} \label{sec:MechanicsModel}

As the condensate grows, the incorporation of new material stretches the scaffold and strains the condensate.
Over time, this strain may relax due to the viscoelastic material response of the scaffold.
We quantify condensate strain by the left Cauchy--Green tensor $\mathbf{B}$. 
Following an Eulerian description of elasticity \citep[\textit{e.g.,}][]{mokbel2018phase, Wei2023}, and assuming that only motion of the scaffold component~$S$ contributes to the accumulation of strain, we arrive at the following dynamic equation for~$\mathbf{B}$,
 \begin{equation} \label{eq:StrainEvolution3D}
  \stackrel{\triangledown}{\mathbf{B}} = -\frac{1}{\tau}\left(\mathbf{B}-\mathbf{I}\right) \;,
 \end{equation}
where $ \stackrel{\triangledown}{\mathbf{B}} =\partial_t\mathbf{B}+\left[\vect{v}\cdot\nabla\right]\mathbf{B}-\left[\nabla\vect{v}\right]^{\intercal}\cdot\mathbf{B}-\mathbf{B}\cdot\left[\nabla\vect{v}\right]$ is the upper-convected derivative, $\mathbf{I}$ is the identity tensor, and $\tau$ is the characteristic relaxation time.
The left side of \eqref{eq:StrainEvolution3D} represents the rate of change of $\mathbf{B}$ with respect to a frame of reference both co-moving and co-rotating with the scaffold, as driven by the scaffold velocity $\vect{v}$, and the right side characterises the relaxation of $\mathbf{B}$ to the undeformed state ($\mathbf{B}=\mathbf{I}$) over timescale $\tau$. At the surface of the active core, $\mathbf{B}$ satisfies a homogeneous Neumann boundary condition.
To calculate the elastic stress $\boldsymbol{\sigma}$ from the strain $\mathbf{B}$, we use a Neo-Hookean model~\cite{Ogden1997}, %
\begin{equation}\label{eq:Stress3D}
\boldsymbol{\sigma}=\KB(J-1)\mathbf{I} +\KS J^{-\frac{5}{3}}\left[\mathbf{B}-\frac{1}{3}\mathrm{tr}\left(\mathbf{B}\right)\right] \; ,
\end{equation}
where $J=\sqrt{\textrm{det}(\mathbf{B})}$ quantifies the relative volume of a material element to its corresponding volume in the undeformed state.
$\KB$ and $\KS$ are the respective bulk and shear moduli of the scaffold. For simplicity, we take $\KB=\KS=K$ throughout.

Finally, flux of scaffold material within the condensate is driven by gradients in both osmotic stress $\boldsymbol{\Pi}$ (due to chemical forces)
and elastic stress $\boldsymbol{\sigma}$ (due to mechanical forces), such that
\begin{equation}\label{eq:Flux3D}
\vect{j}=-\Lambda \nabla  \cdot \left[\boldsymbol{\Pi}- \boldsymbol{\sigma} \right] \;,
\end{equation}
where $\Lambda$ is the mobility of scaffold material.
In general, the osmotic stress is a function of $\phi_S$ that depends on the specific energetic interactions between the scaffold and its pore fluid. For a Flory--Huggins mixing energy, for example, the osmotic stress is isotropic, with diagonal components given by ${\Pi = -k_\mathrm{B}T\left[\phi_S+\log(1-\phi_S)-\chi\phi_S^2\right]}$.
Here, $k_\mathrm{B}T$ is the characteristic energy scale, and $\chi$ is the Flory interaction parameter.
Variations in either $\phi_S$ or $\mathbf{B}$ (and hence $\boldsymbol{\sigma}$) can thus cause a change in the scaffold flux.
In turn, the scaffold flux drives changes in both $\phi_S$ and $\mathbf{B}$ (via \eqref{eq:ContMassBalance} and \eqref{eq:StrainEvolution3D}, respectively), therefore providing a two-way coupling between the reaction--diffusion kinetics discussed above and viscoelastic deformation of the condensate.

\subsection{{Elastic stresses favour a uniform scaffold}}\label{sec:PhaseField}

To illustrate the key features of our system, we first conduct detailed simulations of the above equations by using a phase-field formulation, in which the boundary between the condensate interior and exterior is approximated as a diffuse interface. In this framework, we treat $\phi_S$ as a phase-field order parameter that smoothly interpolates between the value of $\phi_S$ inside the condensate and zero outside the condensate. 
We describe the transport of precursor material in the cell by the reaction--diffusion equation ${\partial_t \phi_P =  \nabla \left[D\cdot\nabla \phi_P\right] - s}$, where $D$ is a diffusion coefficient that may depend on composition.
Along with this evolution equation for $\phi_P$, Eqs.~(\ref{eq:ContMassBalance}-\ref{eq:Flux3D}) give a closed system of equations that can be solved to find $\phi_S(\vect{r},t)$ and $\mathbf{B}(\vect{r},t)$. 
Assuming spherical symmetry, we solve these via the finite element scheme detailed in the \Supp.

\begin{figure}%
    \centering
    \includegraphics[width=\columnwidth]{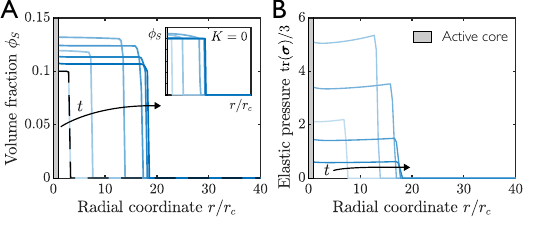}
        \caption{Example phase-field simulation of condensate growth, showing how the spatial distribution of scaffold material (A) and elastic pressure (B) evolve over time (light to dark curves). The inset shows corresponding results for a viscous condensate ($K=0$).
        Results are plotted at evenly spaced times up to $k_-t = 0.4$. The dashed black line shows the initial condition.    
        Reaction parameter values are $k_+/k_-=400$ and $q/k_-\rc=100$, and the rescaled viscoelastic relaxation time is $k_-\tau=0.1$.
        Other parameter values related to the phase-field formulation are provided in the \Supp.
           }
    \label{fig:PhaseField}
\end{figure}

\Figref{fig:PhaseField}A shows how $\phi_S(r)$ evolves over time during condensate growth. %
Initially, scaffold material occupies only a small portion of the total cell volume, located near to the core.
As precursor material is converted into additional scaffold material, the condensate grows. 
To accommodate newly incorporated $S$ components, the existing scaffold is stretched, inducing a tensile stress throughout the condensate (\figref{fig:PhaseField}B).
This elastic stress resists condensate expansion, leading to an increased scaffold fraction $\phi_S$ and smaller condensate radius relative to a purely viscous material.
Over time, stress relaxes, resulting in a reciprocal decrease in $\phi_S$.
Although the mean value of $\phi_S$ in the condensate varies during growth, it remains approximately uniform in space.
In contrast, simulations of a viscous condensate show marked spatial gradients in $\phi_S$ as scaffold fluxes outwards from the centre of the condensate (\figref{fig:PhaseField}A; inset).
This observation suggests that elastic stresses act to promote a spatially uniform scaffold, and motivates the development of a reduced-order model that does not resolve spatial variations in $\phi_S$.

\subsection{Integrated model describes growth dynamics} \label{sec:ReducedModel}

Informed by our phase-field simulations, we now derive a simplified model that explicitly describes the evolution of the condensate radius $R$. %
Integrating \eqref{eq:ContMassBalance} over the volume of the condensate, we find
\begin{equation}\label{eq:GrowthEquation}
    \dd{R}{t}=\frac{R^3-\rc^3}{3 \bar{\phi}_{S}R^2}\left[\bar{s}-\dd{ \bar{\phi}_{S}}{t}\right]+\frac{q\phi_P^\mathrm{c}\rc^2}{ \bar{\phi}_{S} R^2} \; .
 \end{equation}
Here $\bar{\phi}_S$ is the mean value of $\phi_S$ in the condensate, $\phi_P^\mathrm{c}$ is the precursor fraction at the surface of the core, and ${\bar s=3(R^{3}-\rc^3)^{-1}\int_{\rc}^{R}\left(k_{+}{\phi}_{P}-k_{-} \bar{\phi}_{S} \right)r^2\textrm{d}r}$ is the mean incorporation of scaffold material within the bulk of the condensate.
Approximating the scaffold fraction $\phi_S$ by its mean $\bar\phi_S$ is equivalent to assuming that the rearrangement of scaffold material within the condensate as a result of osmotic gradients is fast compared to the rate of condensate growth. %
The radial scaffold flux $j_r$ is then driven only by local material conservation, implying
\begin{equation}\label{eq:jExpression}
    j_r(r)\simeq\frac{r^3-\rc^3}{3r^2}\left[\bar s-\dd{\bar{\phi}_S}{t}\right]+\frac{q\phi_P^\mathrm{c}\rc^2}{r^2} \; .
\end{equation}
A detailed derivation of these equations is provided in the \Supp. 

To derive an expression for $\bar{\phi}_S$, we consider the balance between osmotic pressure and elastic stress. The osmotic pressure acts to set an energetically preferred value of the scaffold volume fraction, denoted by $\phi_S^0$. 
For a porous gel-like material, $\phi_S^0$ is the relaxed (or reference) volume fraction of the scaffold, and for a phase-separated droplet-like material, it corresponds to the equilibrium volume fraction associated with the dense phase.
The actual value of $\bar\phi_S$ may deviate from this preferred value due to the action of internal elastic stresses.
Similarly, external forces, quantified by the radial surface stress $\sigma^*$, may also deform the condensate from its preferred configuration. %
Assuming that $\bar{\phi}_S$ does not deviate significantly from $\phi_S^0$, we thus write
\begin{equation}\label{eq:phiEquation}
     \bar{\phi}_{S}\simeq\phi_{S}^0+\frac{3\alpha}{R^3-\rc^3}\int_{\rc}^R\textrm{tr}(\boldsymbol{\sigma}) r^2\textrm{d}r+\alpha\sigma^*(R) \; ,
\end{equation}
where $\textrm{tr}(\boldsymbol{\sigma})=\sigma_{rr}+2\sigma_{\theta\theta}$ is proportional to the elastic pressure, and $\alpha$ is a small parameter ($\alpha \ll K^{-1}$) that quantifies the influence of osmotic effects.

Finally, we assume that precursor material diffuses sufficiently rapidly that we can approximate $\phi_P$ by its mean value $\bar{\phi}_P$.
Similarly, we take the volume fraction of precursor material outside the condensate to be uniform and given by $\phi_P^{\textrm{out}}=\eta\bar{\phi}_{P}$, where $\eta$ quantifies any differential partitioning of precursor between the condensate interior and exterior.
As a result, $\bar{\phi}_{P}$ can be calculated directly from $\bar{\phi}_{S}$ and the condensate volume $V=\frac43 \pi (R^3-r_\mathrm{c}^3)$ via conservation of total condensate material.
Eqs. (\ref{eq:GrowthEquation}-\ref{eq:phiEquation}), in addition to the radial and azimuthal normal components of \eqref{eq:StrainEvolution3D} and \eqref{eq:Stress3D}, then provide a closed system of equations that describes the evolution of the condensate radius, $R(t)$, and the strain profile within the scaffold, $B_{rr}(r,t)$ and  $B_{\theta\theta}(r,t)$. 
These equations comprise a coupled ODE--PDE system that can be solved numerically given appropriate initial conditions.

\subsection{Bulk incorporation accelerates growth}\label{sec:UncoupledGrowth}

\begin{figure}
    \centering
    \includegraphics[width=\columnwidth]{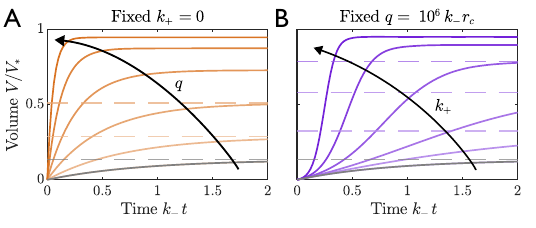}
        \caption{Growth curves for different reaction schemes, showing rescaled volume $V/V_*$ as a function of rescaled time $k_-t$, where $V_*=\bar{\phi}\Vc/\phi_S^0$ is the maximum attainable condensate volume. (A) Effect of increasing core incorporation $q$ at fixed bulk reaction rate $k_+=0$. Light to dark curves show $q/k_-\rc$ ranging from $10^6$ to $10^8$. (B) Effect of increasing $k_+$ at fixed $q/k_-\rc=10^6$. Light to dark curves show $k_+/k_-$ ranging between zero and $10^5$. The grey curves in each plot show results for the same sets of reaction rates. 
        Dashed lines show the corresponding final volume for simulations which do not reach their steady size until $k_-t>2$.
       Other parameter values are $\phi_S^0=0.1$, $\bar{\phi}=2\times10^{-4}$, $\eta=1$, $\Vc/\rc^3=5\times10^8$ and $R_0/\rc=2$. %
        }
    \label{fig:UncoupledGrowth}
\end{figure}

To focus on the underlying growth kinetics, and to understand the role of the two different incorporation channels (bulk and core-localised), we first neglect the feedback of elastic stress on scaffold density, setting $\alpha=0$ such that $\bar{\phi}_S=\phi_S^0$ for all time.
\Figref{fig:UncoupledGrowth} shows the evolution of condensate volume over time for different reaction schemes in this case.
In general, condensate growth results from the conversion of precursor material into scaffold material within the condensate.
During growth, precursor material is thus depleted, until eventually a steady final volume, denoted by $\Vf$, is reached.
In the absence of elastic feedback, $\Vf$ is set by a balance between the rate of scaffold incorporation, the turnover of scaffold material back into precursor, and the mean volume fraction of condensate material in the cell.
For all reaction schemes, $V$ shows exponential saturation as it approaches $\Vf$.
At early times, core-localised reactions lead to linear growth  ($k_+=0$; \figref{fig:UncoupledGrowth}A). In contrast, including bulk reactions gives sigmoidal growth ($k_+>0$; \figref{fig:UncoupledGrowth}B).
Increasing the reaction parameters $q$ and $k_+$ results in faster growth and a larger final condensate for both reactions schemes.
Although core-localised incorporation is essential to ensure initial condensate formation~\cite{Zwicker2014}, extremely large values of $q$ are required to achieve similar growth rates and condensate sizes to those achieved when bulk incorporation is also included.
As condensates grow larger, the availability of precursor material at the core to fuel core incorporation may also be limited by the finite diffusion time of precursor through the condensate~\cite{Zwicker2014}.
Further, high rates of core incorporation are in fact inhibited by the large strains that are typically induced close to the core (as discussed below).
As such, the inclusion of a bulk incorporation mechanism allows for an acceleration of condensate growth, and for cells to readily grow condensates that are many times larger than the active core.

\subsection{{Scaffold growth induces large strains near the core}}\label{sec:UncoupledMechanics}

\begin{figure}
    \centering
    \includegraphics[width=\columnwidth]{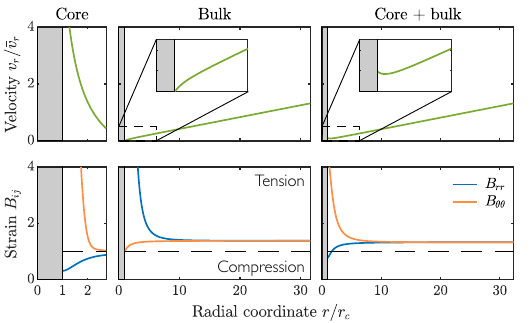}
    \caption{Variation of scaffold velocity (top) and strain (bottom) as functions of radial coordinate $r/\rc$ for different reaction schemes. 
    Insets highlight the velocity profile near to the core (solid grey region).
Blue lines show radial normal strain $B_{rr}$, and orange lines azimuthal normal strain $B_{\theta\theta}$. The dashed black lines at $B_{ij}=1$ correspond to no deformation. 
This snapshot is taken at $k_-t=0.5$. We use the following reaction parameters: $k_+/k_-=0$, $q/k_-\rc=5\times 10^3$ (left); $k_+/k_-=10^4$, $q/k_-\rc=0$ (centre); $k_+/k_-=10^4$, $q/k_-\rc=6\times 10^3$ (right); and set $k_-\tau=0.04$. All other parameter values are the same as in \figref{fig:UncoupledGrowth}.
} 
    \label{fig:Mechanics}
\end{figure}

We now study the mechanics of condensate growth, highlighting how different modes of material incorporation induce qualitatively different states of deformation.
In general, incorporation of new scaffold components drives an outward flux of material, quantified by $j_r$, that strains the existing scaffold network.
In the absence of bulk incorporation ($k_+=0$), all new material is inserted at the surface of the active core and fluxes outwards from the condensate centre.
Consequently, $j_r$ is largest near to the core, and decreases monotonically towards the outer boundary of the condensate (\figref{fig:Mechanics}; left).
This flux induces a radial compression ($B_{rr}<1$) and azimuthal tension ($B_{\theta\theta}>1$) in the scaffold, which reduce with distance $r$ from the centre of the condensate.
In contrast, bulk reactions promote isotropic material incorporation, such that new scaffold components are inserted throughout the condensate volume, and $j_r$ increases almost linearly with $r$. 
The scaffold flux at the surface of the core $r=\rc$ is set by the rate of core incorporation. If $q=0$, then $j_r=0$ at $r=\rc$  (\figref{fig:Mechanics}; centre).
In this case, scaffold molecules initially situated at the core surface remain at the core surface during condensate growth, and so cannot be stretched in the azimuthal direction ($B_{\theta\theta}|_{r=\rc}=1$).
In the radial direction, however, the outwards flux of condensate material stretches the scaffold and induces large tensile strains.
The scaffold flux resulting from the combination of both core and bulk incorporation is as expected; $j_r$ increases linearly far away from the core, but is non-zero at the core surface (\figref{fig:Mechanics}; right). Correspondingly, we find large strains near to the core (which may be either tensile or compressive depending on the magnitude of $q$), and uniform tensile strains away from the core.

For all reaction schemes, strain continues to increase in magnitude as the condensate grows, until $t\sim\tau$ such that viscoelastic relaxation becomes significant.
We also find that strain (and hence elastic stress) is always largest near the core. 
In practice, extremely large strains will prohibit further incorporation of new material, and so very high rates of core incorporation will not be possible.
As such, we now focus on systems in which bulk incorporation is the dominant driver of growth, and seek to understand how internal elastic stress affects condensate growth dynamics.

\subsection{Internal elasticity restricts condensate growth}\label{sec:CoupledGrowth}

\begin{figure}%
    \centering
    \includegraphics[width=\columnwidth]{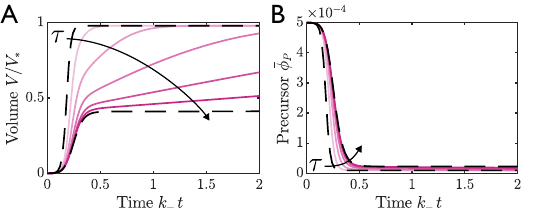}
    \caption{The effect of viscoelastic stresses on condensate growth, showing rescaled volume $V/V_*$ (A) and precursor fraction $\bar{\phi}_{P}$ (B) as functions of rescaled time $k_-t$. Light to dark curves show results for different values of the relaxation time, ranging from $k_-\tau=0.05$ to $k_-\tau=2$. The black dashed lines show the limits $\tau\to0$ and $\tau\to\infty$. Parameter values are $k_+/k_-=10^4$, $q/k_-\rc=0.01$, $\alpha K=0.002$, and $\sigma^*=0$. Other parameter values are the same as in \figref{fig:UncoupledGrowth}.
    }
    \label{fig:CoupledGrowth}
\end{figure}

To investigate the impact of internal elasticity on condensate growth, we now couple scaffold density to elastic deformation, allowing $\bar{\phi}_S$ to vary with stress as described by \eqref{eq:phiEquation}.
\Figref{fig:CoupledGrowth}A shows that the influence of elastic stress on growth depends strongly on the rate of stress relaxation, quantified by $\tau$, compared to the rate of material incorporation.
In the limit $k_+ \ll \tau^{-1}$, strains relax sufficiently fast that condensate growth is unaffected by internal stresses.
However, for $k_+ \gg \tau^{-1}$ strain relaxation is slow compared to the rate of material incorporation and growth is significantly inhibited.
Here, we keep $k_+$ and $q$ constant, implying that new material is incorporated into the condensate at the same rate regardless of the stress state.
Consequently, elastic stresses have only a small effect on the rate of precursor material consumption (\figref{fig:CoupledGrowth}B).
This result highlights that the restricted growth is driven by increased deformation of the scaffold, rather than by reaction kinetics.
We explain the impact of elasticity on growth as follows: as new material is incorporated into the condensate, the scaffold volume fraction $\phi_S$, and hence the osmotic pressure, increase locally. 
To reduce osmotic pressure and obtain the preferred volume fraction $\phi_S^0$, the condensate expands, with solvent swelling the scaffold. 
This expansion stretches the scaffold, inducing a large tensile stress. This stress then acts against the osmotic pressure, resisting the increase in tension associated with further swelling.
As a result, the condensate cannot expand fully, and remains at a smaller size, with $\bar{\phi}_S>\phi_S^0$.
Over time, stress relaxation allows the osmotic pressure to drive further scaffold expansion, until eventually $\bar{\phi}_S=\phi_S^0$ and the condensate reaches its final (unstrained) volume.
Faster stress relaxation (corresponding to a more liquid-like rheology) therefore facilitates faster condensate growth.

\subsection{{Viscoelastic material properties impart mechanical strength}}\label{sec:Deformation}

We now characterise condensate mechanical strength by applying a constant radial stress $\sigma^*$ to a fully grown condensate (reactions complete) with volume $V_0$.
In general, this stress will deform the condensate by stretching the scaffold. We quantify condensate deformation by the normalised volumetric expansion $\epsilon = (V-V_0)/V_0$.
Typically, there exists a critical stretch associated with material failure that we denote by $\VCrit$.
If $\epsilon > \VCrit$, the condensate will rupture and no longer be able to anchor forces.
For a liquid-like condensate (small $\tau$), the external stress is only resisted by relatively weak osmotic and surface tension effects, and so the scaffold is easily deformed.
In this case, the stress needed to initiate material failure is given by $\alpha\sigma^* = {\VCrit\phi_S^0}/({1+\VCrit})$ (\Supp).
In contrast, viscoelastic condensates are reinforced by internal elastic stresses, and hence resist deformation much more strongly.
For a completely solid-like rheology (large $\tau$), material failure does not occur until $\alpha\sigma^* = {\VCrit\phi_S^0}/({1+\VCrit})+\alpha K\VCrit  $.
As an example, a liquid-like condensate with $\VCrit=10$ and $\phi_S^0=0.1$ can anchor a maximum stress $\alpha\sigma^*\sim0.1$.
Alternatively, a solid-like condensate also with $\alpha K=1$ can anchor a maximum stress $\alpha\sigma^*\sim10$, representing a 100-fold increase in strength.
For intermediate relaxation time~$\tau$, condensates initially resist deformation in an elastic-like manner, but then continue to expand over time as internal stresses relax.
The time scale over which condensates must resist deformation depends on their specific function.
To determine whether it is possible for a viscoelastic condensate to display both rapid growth and appropriate mechanical strength, we thus apply our theoretical framework to the specific case of centrosomes in \textit{C. elegans} embryos, identifying constraints on material parameters that can reconcile these requirements.

\section{Centrosomes as growing viscoelastic materials}\label{sec:Centrosomes}

Centrosomes are membrane-less organelles that act as the main microtubule organising centre in metazoans.
Centrosomes consist of a well-defined core (a pair of centrioles), which is surrounded by an amorphous, mesh-like structure known as the pericentriolar material (PCM).
The PCM comprises a porous scaffold network permeated by both solvent and client molecules~\cite{Schnackenberg1998}. 
Recent experiments have shown that the PCM in \textit{C. elegans} centrosomes exhibits a characteristic pore size~\cite{Tollervey2025}, and displays a viscoelastic rheology~\cite{Amato2024}.
During mitosis, centrosomes mature rapidly, with the PCM dramatically increasing in size.
Incorporation of new scaffold material is driven by phosphorylation-dependent conversion of precursor molecules into an assembly-competent form within the pre-existing PCM~\cite{Wueseke2016,Nakajo2022,Rios2024}.
As centrosomes grow, microtubules nucleate within the PCM, extending outwards into the surrounding cell. %
These microtubules utilise centrosomes as mechanical anchor points for transducing mechanical forces that organise the mitotic spindle~\cite{Young2025}.
Current models of centrosome structure, however, are unable to reconcile the twin requirements of rapid (fluid-like) PCM assembly and (solid-like) mechanical strength~\cite{Raff2019,Woodruff2021}. %
To demonstrate the ability of our modelling framework to explain this dichotomy, we compare the predictions of our model to experimental data for centrosomes in \textit{C. elegans} embryos, following the procedure detailed in Ref.~\cite{GarciaBaucells2025}.

In our theoretical framework, the assembly-competent component $S$ takes the role of the PCM scaffold, and the active core that of the centrioles.
The action of microtubule-mediated forces is represented by the external stress $\sigma^*$.
We emphasise that our description of centrosome structure is agnostic as to whether the PCM microstructure constitutes a gel-like lattice structure or a phase-separated droplet-like material.

\begin{figure*}
    \centering
    \includegraphics[width=0.9\textwidth]{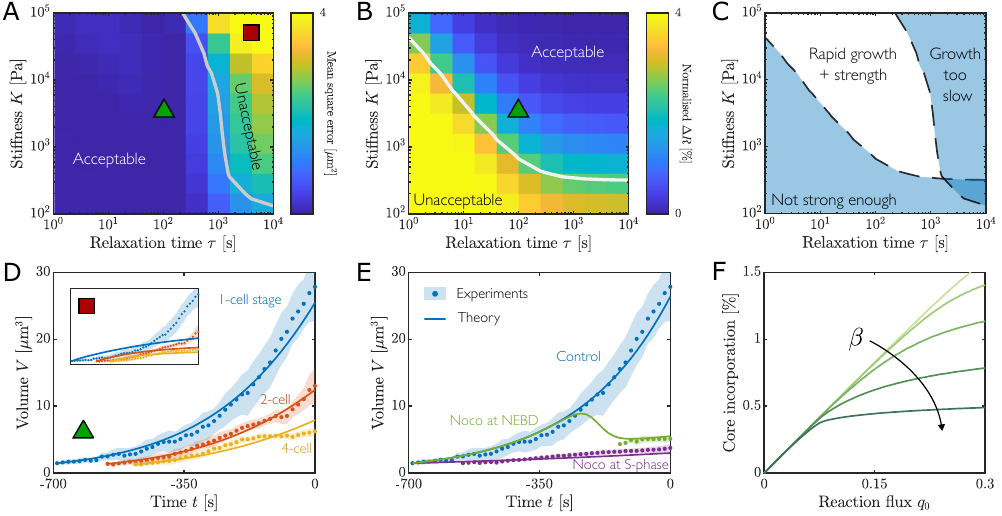}
\caption{(A) Phase space showing the quality of fit between simulation results and experimental growth curves for different elastic modulus $K$ and relaxation time $\tau$. The white curve separates parameter values for which an acceptable fit is possible, and for which it is not. 
 (B) Normalised change in centrosome radius $\Delta R$ induced by a mitotic stress for various $K$ and $\tau$. The white curve separates parameter values for which scaffold expansion is acceptable, and for which it is not.
(C) Combining results from panels (A) and (B) provides constraints on the material properties that allow both rapid growth and mechanical strength.
(D) Comparison of model predictions (solid lines) and experimental data (scattered dots) for centrosome growth at various cell stages, assessed using GFP:SPD-5. For the experiments, $n=17$ embryos (one-cell stage), $18$ (two-cell), and $15$ (four-cell). Time is measured relative to cytokinesis onset of each division cycle.
Values of material parameters indicated by green triangle in panel (A). 
Inset shows model predictions in the ‘unacceptable’ parameter regime (red square). 
(E) Depolymerisation of microtubules via nocodazole during both S-phase (purple) and at NEBD (green) inhibits scaffold growth compared to the control case (blue). Experiments were performed for control embryos ($n = 14$ embryos), as well as embryos treated with nocodazole at S-phase ($n= 9$) and at NEBD ($n = 13$) of the first embryonic mitosis. Note slight reduction in PCM size upon extended treatment with nocodazole in line with model predictions.
Time is measured relative to chromosome decondensation.
(F) Proportion of total scaffold material that is incorporated at the centriole as a function of base incorporation rate~$q_0$. Light to dark lines show simulations for increasing $\beta$ from $\beta=10^{-5}$ to $\beta=0.1$, which characterises how strongly reactions depend on strain.
Other parameter values are the same as in \figref{fig:CoupledGrowth}. 
In panels (D) and (E), scattered dots show mean experimental values, and shaded areas show one standard deviation about the mean.
We convert our results to dimensional units by setting $k_- = 10^{-3}\,\textrm{s}$, $\rc = 75 \, \textrm{nm}$, and $\alpha = 0.1\, \textrm{kPa}^{-1}$.
    }
    \label{fig:Experiments}
\end{figure*}

\subsection{{}{Liquid-like centrosomes permit rapid scaffold growth}}

During centrosome maturation, the PCM scaffold undergoes rapid expansion as new scaffold proteins (\textit{e.g.},~SPD-5 in the case of \textit{C. elegans}) are incorporated. 
However, scaffold incorporation also drives large strains within the PCM, which can restrict growth (\figref{fig:CoupledGrowth}).
Comparing experimental growth curves for centrosomes from embryos at the 1, 2 and 4 cell stages with model predictions, we can constrain the mechanical parameters that allow sufficiently rapid growth (\figref{fig:Experiments}A).
To determine permissible material properties, we fit our model to experimental data by varying $k_+$ and $\bar\phi$ for given values of the relaxation time $\tau$ and elastic modulus $K$. 
We keep the rate of centriolar incorporation $q$ constant throughout the fit, since it has a very limited effect on the overall growth curves.
Quantifying the goodness of fit for each combination of material parameters by the mean squared error (\figref{fig:Experiments}A), we then identify `acceptable' and `unacceptable' regions of the parameter space for $\tau$ and $K$ spanning several orders of magnitude.
We note that because elastic effects enter our model in combination with the phenomenological parameter $\alpha$, the dimensional values of $K$ presented here can be rescaled without affecting our general conclusions.
For large values of $\tau$ and $K$ (solid-like rheology), we find that model predictions are incompatible with experimental data.
In this parameter regime, a good fit between experiment and theory is not possible (\figref{fig:Experiments}D; inset), regardless of the fit parameters $k_+$ and $\bar{\phi}$, since elastic stresses limit the maximum rate of growth.
On the other hand, for small values of $\tau$ and $K$ (liquid-like rheology), an excellent fit is possible (\figref{fig:Experiments}D; main).

\subsection{{Viscoelastic centrosomes can sustain mitotic forces}}

As well as exhibiting rapid growth during maturation, centrosomes must anchor large mechanical forces during mitosis. 
The action of mitotic forces is represented in our model by the external stress $\sigma^*$. 
Applying a mitotic force of $100\,\textrm{pN}$~\cite{GarzonCoral2016,Mittasch2020} to a fully grown centrosome, we then simulate the subsequent volumetric expansion of the scaffold over a time period of $100\,\textrm{s}$.
For relatively liquid-like material properties (small $\tau$, $K$), we observe significant expansion of the scaffold, as quantified by the normalised change in radius $\Delta R$ (\figref{fig:Experiments}B).
Such a large deformation may result in material failure and disassembly of the PCM.
For more solid-like properties (large $\tau$, $K$), however, the scaffold is able to resist deformation and only expand by a small amount.
As before, we now identify regions of the parameter space for which the simulated expansion is acceptable and for which it is not, using an expansion threshold of $1\%$.
Combining the constraints identified in \figsref{fig:Experiments}A-B, we can thus identify a parameter regime for which both rapid growth and mechanical strength are possible (\figref{fig:Experiments}C).

\subsection{Microtubule-mediated stress drives scaffold expansion}

To assess the impact of microtubule-mediated stresses on centrosome growth during maturation, we conducted experiments in which embryonic \textit{C. elegans} is treated with nocodazole, which acts to depolymerise microtubules throughout the cell.
This process can be simulated in our modelling framework by setting $\sigma^*=0$ at the point of nocodazole addition.
Due to the finite time of nocodazole permeation, we enforce this change in $\sigma^*$ smoothly over a time period of $30\,\textrm{s}$. %

\Figref{fig:Experiments}E shows centrosome volume as a function of time (measured relative to chromosome decondensation) for embryos treated with nocodazole both at nuclear envelope breakdown (NEBD) and during S-phase.
For both treatment protocols, the final centrosome volume is significantly reduced in comparison to the control case.
For centrosomes treated at NEBD, growth before treatment follows that of the control case. Once nocodazole is added, however, there is a sudden drop in centrosome volume. Notably, this decrease in volume does not coincide with a decrease in the total amount of scaffold protein in the PCM but instead results from an increase in scaffold density~\cite{GarciaBaucells2025}.
From a fit of model predictions with $\tau=100\, \textrm{s}$ to the experimental growth curves, we estimate the ratio of microtubule-mediated stress~$\sigma^*$ to the elastic modulus~$K$, which is on the order of $\sigma^*/K\sim30 $.
For centrosomes treated during S-phase (\textit{i.e.}, before maturation has begun), centrosome volume is always smaller than for the control case. Additionally, the final centrosome volume is also smaller than for centrosomes treated during NEBD.
We hypothesise that this result is due to NEBD-treated centrosomes initially growing under the expansion of the microtubule-mediated stress, thus leading to more bulk incorporation.

\subsection{Stress-dependent reactions promote bulk incorporation}

If the PCM is significantly strained, it may become harder to incorporate new material, because new scaffold molecules may have to deform to attach to available binding sites.
Our model can incorporate such an effect by allowing the reaction parameters $k_+$ and $q$ to be functions of stress (or equivalently strain).
We test the impact of such a dependence on scaffold incorporation by considering a simple form of the centriolar incorporation rate, ${q=q_0\exp\left[-\beta|\textrm{tr}(\boldsymbol{\sigma})|\right]}$, where $\beta>0$ characterises the strength of the stress dependence. We take an equivalent form for the bulk incorporation rate $k_+$, with the same value of $\beta$. This choice implies that incorporation rates are reduced for both tensile and compressive stresses.

As highlighted in \figref{fig:Mechanics}, strain is always largest in the vicinity of the active core (centriole).
As such, stress-dependent reaction rates inhibit material incorporation more strongly near to the centriole than in the rest of the PCM.
To quantify this effect, we calculate the total proportion of all scaffold material that is incorporated at the centriole compared to the rest of the PCM over the course of centrosome growth (\figref{fig:Experiments}F).
For constant incorporation rates ($\beta=0$; $k_+$, $q$ constant), the proportion of scaffold material that is incorporated at the centriole increases linearly with the base reaction rate $q_0$.
For $\beta>0$, however, centriolar incorporation is increasingly unfavourable compared to incorporation away from the centriole.
As such, the maximum proportion of scaffold material that can be incorporated at the centriole is capped, despite arbitrarily large $q_0$.
As a result, stress-dependent reactions tend to promote bulk, rather than centriolar, incorporation.

\section{Conclusions}

The precise and controlled assembly of mechanical anchor points is vital to the function of a range of biological and synthetic materials, including biomolecular condensates.
In this manuscript, we have developed a continuum model that demonstrates how viscoelastic material properties, coupled with actively driven incorporation of new material around a central core, can facilitate the formation of such structures.
Our model thus provides a theoretical framework with which to analyse the growth and deformation of these objects, and to constrain the material properties that are consistent with experimental observations.
Although viscoelastic stresses restrict condensate growth, we find that it is possible to combine both rapid growth and mechanical strength in an appropriate parameter regime.
We also demonstrate that localised incorporation of new condensate material in the vicinity of the active core may be suppressed if incorporation rates depend on condensate stress state. 
Finally, we have demonstrated the applicability of our theory by comparing our model predictions to experimental data for \textit{C.~elegans} centrosomes~\cite{GarciaBaucells2025}, quantitatively capturing the growth dynamics of both wild-type and nocodazole-treated samples.
Taken together, our findings suggest that active viscoelastic condensates provide a simple solution for cells to rapidly form mechanical anchor points.

The application of our modelling framework to the case of centrosomes represents a significant development on previous theories of centrosome growth~\citep[\textit{e.g.}][]{Zwicker2014,Banerjee2025}, reconciling the fact that centrosomes must display suitable mechanical strength in addition to rapid maturation, and identifying a parameter regime in which this is possible.
The broad range of allowable elastic moduli suggests that this mechanism is robust even allowing for recently observed softening of the PCM in the approach to mitosis~\cite{GarciaBaucells2025}.
In addition, we have shown that elastic stresses may inhibit the incorporation of new scaffold material, particularly in the vicinity of the centriole, where stresses are largest.
This effect could counteract the expected enhanced incorporation resulting from a diffusible signal emanating from the centriole~\cite{Cabral2019}, potentially explaining previously observed isotropic incorporation~\cite{Laos2015}.
In summary, our model can explain many aspects of centrosome formation and growth, and lays a key foundation for developing future understanding.

In this work, we have made several simplifying approximations to focus on the core physical principles that govern the formation, growth and deformation of viscoelastic condensates in cells.
Relaxing these assumptions will facilitate detailed investigation of the effects of physical phenomena not considered here. 
For example, we have focussed primarily on the limiting case in which precursor material distributes rapidly and uniformly throughout the cell.
Our integrated model can be straightforwardly extended to situations in which the finite diffusion time of precursor is important by explicitly tracking the diffusion-governed motion of precursor material, as we already do in our phase-field simulations.
Further, the influence of time-dependent material properties \citep[\textit{e.g.},][]{Mittasch2020, Jawerth2020,Takaki2023,GarciaBaucells2025}, could be trivially incorporated into our model given a functional dependence of material stiffness or relaxation time on condensate age.
Lastly, although our integrated model neglects variations in the spatial distribution of scaffold material inside a condensate, the impact of gradients in scaffold deformation (\textit{e.g.}, due to poroelastic drag) can be readily explored via the detailed phase-field simulations presented in \figref{fig:PhaseField}.
As such, we expect that the general framework developed here will be useful for studying a wide range of systems, including other examples of biomolecular condensates (\textit{e.g.}, focal adhesions and tight junctions), as well as tissues at the multi-cellular scale, and biomimetic materials such as bio-adhesives and tissue engineering scaffolds~\cite{Harrington2024}.

\section*{Acknowledgements}

The authors thank Sebastian Fürthauer for helpful discussions.
OWP and DZ acknowledge funding from the Max Planck Society and the European Union (ERC, EmulSim, 101044662).
This work was also supported by funding from the Austrian Science Fund FWF, grant P34526-B, to AD, as well as a Max Perutz PhD fellowship of the Max Perutz Labs to JG-B.
SA and LZ acknowledge support from DFG (grant 511509575).
SA and LZ also gratefully acknowledge computing time on the high-performance computer at the NHR Center of TU Dresden. This centre is jointly supported by the Federal Ministry of Education and Research and the state governments participating in the NHR (www.nhr-verein.de/unsere-partner).
This research was supported in part by grant NSF PHY-2309135 and the Gordon and Betty Moore Foundation Grant No. 2919.02 to the Kavli Institute for Theoretical Physics (KITP). 

\section*{Author contributions}

OWP and DZ designed research;
OWP, LZ, SA, and DZ developed the model;
OWP and LZ performed simulations;
JG-B and AD designed and performed experiments;
OWP analysed data;
OWP wrote the original draft; 
and all authors edited the final manuscript.

\bibliographystyle{unsrtnat}
\bibliography{viscoelastic_centrosomes.bib}

\begin{thebibliography}{43}
\providecommand{\natexlab}[1]{#1}
\providecommand{\url}[1]{\texttt{#1}}
\expandafter\ifx\csname urlstyle\endcsname\relax
  \providecommand{\doi}[1]{doi: #1}\else
  \providecommand{\doi}{doi: \begingroup \urlstyle{rm}\Url}\fi

\bibitem[Bracha et~al.(2019)Bracha, Walls, and Brangwynne]{Bracha2019}
Dan Bracha, Mackenzie~T Walls, and Clifford~P Brangwynne.
\newblock Probing and engineering liquid-phase organelles.
\newblock \emph{Nature biotechnology}, 37\penalty0 (12):\penalty0 1435--1445,
  2019.

\bibitem[O'Flynn and Mittag(2021)]{Oflynn2021}
Brian~G. O'Flynn and Tanja Mittag.
\newblock The role of liquid--liquid phase separation in regulating enzyme
  activity.
\newblock \emph{Curr. Opin. Cell Biol.}, 69:\penalty0 70--79, 2021.
\newblock ISSN 0955-0674.

\bibitem[Sanfeliu-Cerd{\'a}n and Krieg(2025)]{Sanfeliu2025}
Neus Sanfeliu-Cerd{\'a}n and Michael Krieg.
\newblock The mechanobiology of biomolecular condensates.
\newblock \emph{Biophysics Reviews}, 6\penalty0 (1):\penalty0 011310, 03 2025.
\newblock ISSN 2688-4089.
\newblock \doi{10.1063/5.0236610}.
\newblock URL \url{https://doi.org/10.1063/5.0236610}.

\bibitem[Brangwynne et~al.(2009)Brangwynne, Eckmann, Courson, Rybarska, Hoege,
  Gharakhani, J{\"u}licher, and Hyman]{Brangwynne2009}
Clifford~P Brangwynne, Christian~R Eckmann, David~S Courson, Agata Rybarska,
  Carsten Hoege, J{\"o}bin Gharakhani, Frank J{\"u}licher, and Anthony~A Hyman.
\newblock Germline p granules are liquid droplets that localize by controlled
  dissolution/condensation.
\newblock \emph{Science}, 324\penalty0 (5935):\penalty0 1729--1732, 2009.
\newblock \doi{10.1126/science.1172046}.

\bibitem[Banani et~al.(2017)Banani, Lee, Hyman, and Rosen]{Banani2017}
Salman~F Banani, Hyun~O Lee, Anthony~A Hyman, and Michael~K Rosen.
\newblock Biomolecular condensates: organizers of cellular biochemistry.
\newblock \emph{Nat. Rev. Mol. Cell Biol.}, 18\penalty0 (5):\penalty0 285--298,
  05 2017.
\newblock \doi{10.1038/nrm.2017.7}.

\bibitem[Dignon et~al.(2020)Dignon, Best, and Mittal]{Dignon2020}
Gregory~L. Dignon, Robert~B. Best, and Jeetain Mittal.
\newblock Biomolecular phase separation: From molecular driving forces to
  macroscopic properties.
\newblock \emph{Annual Review of Physical Chemistry}, 71\penalty0 (Volume 71,
  2020):\penalty0 53--75, 2020.

\bibitem[Yoo et~al.(2019)Yoo, Triandafillou, and Drummond]{Yoo2019}
Haneul Yoo, Catherine Triandafillou, and D~Allan Drummond.
\newblock Cellular sensing by phase separation: Using the process, not just the
  products.
\newblock \emph{Journal of Biological Chemistry}, 294\penalty0 (18):\penalty0
  7151--7159, 2019.

\bibitem[Korkmazhan et~al.(2021)Korkmazhan, Tompa, and Dunn]{Korkmazhan2021}
Elgin Korkmazhan, Peter Tompa, and Alexander~R Dunn.
\newblock The role of ordered cooperative assembly in biomolecular condensates.
\newblock \emph{Nature Reviews Molecular Cell Biology}, 22\penalty0
  (10):\penalty0 647--648, 2021.

\bibitem[Raff(2019)]{Raff2019}
Jordan~W. Raff.
\newblock Phase separation and the centrosome: A fait accompli?
\newblock \emph{Trends in Cell Biology}, 29:\penalty0 612--622, 8 2019.
\newblock ISSN 18793088.
\newblock \doi{10.1016/j.tcb.2019.04.001}.

\bibitem[Woodruff(2021)]{Woodruff2021}
Jeffrey~B. Woodruff.
\newblock The material state of centrosomes: lattice, liquid, or gel?
\newblock \emph{Current Opinion in Structural Biology}, 66:\penalty0 139--147,
  2 2021.
\newblock \doi{10.1016/j.sbi.2020.10.001}.

\bibitem[Litschel et~al.(2024)Litschel, Kelley, Cheng, Babl, Mizuno, Case, and
  Schwille]{Litschel2024}
Thomas Litschel, Charlotte~F. Kelley, Xiaohang Cheng, Leon Babl, Naoko Mizuno,
  Lindsay~B. Case, and Petra Schwille.
\newblock Membrane-induced {2D} phase separation of the focal adhesion protein
  talin.
\newblock \emph{Nature Communications}, 15\penalty0 (1):\penalty0 4986, 2024.
\newblock \doi{10.1038/s41467-024-49222-z}.

\bibitem[Schwayer et~al.(2019)Schwayer, Shamipour, Pranjic-Ferscha, Schauer,
  Balda, Tada, Matter, and Heisenberg]{Schwayer2019}
Cornelia Schwayer, Shayan Shamipour, Kornelija Pranjic-Ferscha, Alexandra
  Schauer, Maria Balda, Masazumi Tada, Karl Matter, and Carl-Philipp
  Heisenberg.
\newblock Mechanosensation of tight junctions depends on zo-1 phase separation
  and flow.
\newblock \emph{Cell}, 179\penalty0 (4):\penalty0 937--952.e18, 2019.
\newblock ISSN 0092-8674.
\newblock \doi{10.1016/j.cell.2019.10.006}.

\bibitem[Beutel et~al.(2019)Beutel, Maraspini, Pombo-Garc{\'\i}a,
  Martin-Lemaitre, and Honigmann]{Beutel2019}
Oliver Beutel, Riccardo Maraspini, Karina Pombo-Garc{\'\i}a, C{\'e}cilie
  Martin-Lemaitre, and Alf Honigmann.
\newblock Phase separation of zonula occludens proteins drives formation of
  tight junctions.
\newblock \emph{Cell}, 179\penalty0 (4):\penalty0 923--936.e11, 2019.
\newblock ISSN 0092-8674.
\newblock \doi{10.1016/j.cell.2019.10.011}.

\bibitem[Sun et~al.(2024)Sun, Zhao, Wiegand, Martin-Lemaitre, Borianne,
  Kleinschmidt, Grill, Hyman, Weber, and Honigmann]{Sun2024}
Daxiao Sun, Xueping Zhao, Tina Wiegand, Cecilie Martin-Lemaitre, Tom Borianne,
  Lennart Kleinschmidt, Stephan~W. Grill, Anthony~A. Hyman, Christoph Weber,
  and Alf Honigmann.
\newblock Assembly of tight junction belts by zo1 surface condensation and
  local actin polymerization.
\newblock \emph{Developmental Cell}, 60\penalty0 (8):\penalty0 1234--1250.e6,
  2024.
\newblock ISSN 1534-5807.

\bibitem[Weber et~al.(2019)Weber, Zwicker, J{\"u}licher, and Lee]{Weber2019}
Christoph~A. Weber, David Zwicker, Frank J{\"u}licher, and Chiu~Fan Lee.
\newblock Physics of active emulsions.
\newblock \emph{Rep. Prog. Phys.}, 82:\penalty0 064601, 2019.

\bibitem[Zwicker et~al.(2025)Zwicker, Paulin, and ter Burg]{Zwicker2025}
David Zwicker, Oliver~W. Paulin, and Cathelijne ter Burg.
\newblock Physics of droplet regulation in biological cells.
\newblock \emph{arXiv}, 2025.
\newblock URL \url{https://arxiv.org/abs/2501.13639}.

\bibitem[Zwicker et~al.(2014)Zwicker, Decker, Jaensch, Hyman, and
  J{\"u}licher]{Zwicker2014}
David Zwicker, Markus Decker, Steffen Jaensch, Anthony~A. Hyman, and Frank
  J{\"u}licher.
\newblock Centrosomes are autocatalytic droplets of pericentriolar material
  organized by centrioles.
\newblock \emph{Proceedings of the National Academy of Sciences}, 111\penalty0
  (26):\penalty0 E2636--E2645, 2014.
\newblock \doi{10.1073/pnas.1404855111}.
\newblock URL \url{https://www.pnas.org/doi/abs/10.1073/pnas.1404855111}.

\bibitem[Mokbel et~al.(2018)Mokbel, Abels, and Aland]{mokbel2018phase}
Dominic Mokbel, Helmut Abels, and Sebastian Aland.
\newblock A phase-field model for fluid--structure interaction.
\newblock \emph{Journal of Computational Physics}, 372:\penalty0 823--840,
  2018.
\newblock ISSN 0021-9991.
\newblock \doi{https://doi.org/10.1016/j.jcp.2018.06.063}.
\newblock URL
  \url{https://www.sciencedirect.com/science/article/pii/S0021999118304418}.

\bibitem[Wei and Wu(2023)]{Wei2023}
Chaozhen Wei and Min Wu.
\newblock An eulerian nonlinear elastic model for compressible and fluidic
  tissue with radially symmetric growth.
\newblock \emph{SIAM Journal on Applied Mathematics}, 83:\penalty0 254--275,
  2023.
\newblock ISSN 00361399.

\bibitem[Ogden(1997)]{Ogden1997}
R.W. Ogden.
\newblock \emph{Non-linear Elastic Deformations}.
\newblock Dover Civil and Mechanical Engineering. Dover Publications, 1997.
\newblock ISBN 9780486696485.

\bibitem[Schnackenberg et~al.(1998)Schnackenberg, Khodjakov, Rieder, and
  Palazzo]{Schnackenberg1998}
Bradley~J Schnackenberg, Alexey Khodjakov, Conly~L Rieder, and Robert~E
  Palazzo.
\newblock The disassembly and reassembly of functional centrosomes in vitro.
\newblock \emph{Cell Biology}, 95:\penalty0 9295--9300, 1998.

\bibitem[Tollervey et~al.(2025)Tollervey, Rios, Zagoriy, Woodruff, and
  Mahamid]{Tollervey2025}
Fergus Tollervey, Manolo~U. Rios, Evgenia Zagoriy, Jeffrey~B. Woodruff, and
  Julia Mahamid.
\newblock Molecular architectures of centrosomes in c. elegans embryos
  visualized by cryo-electron tomography.
\newblock \emph{Developmental Cell}, 60\penalty0 (6):\penalty0 885--900.e5,
  2025.
\newblock ISSN 1534-5807.

\bibitem[Amato et~al.(2025)Amato, Hwang, Rios, Familiari, Rosen, and
  Woodruff]{Amato2024}
Matthew Amato, June~Ho Hwang, Manolo~U. Rios, Nicole~E. Familiari, Michael~K.
  Rosen, and Jeffrey~B. Woodruff.
\newblock Polo-like kinase 1 phosphorylation tunes the functional viscoelastic
  properties of the centrosome scaffold.
\newblock \emph{bioRxiv}, 2025.
\newblock \doi{10.1101/2024.08.29.610374}.
\newblock URL
  \url{https://www.biorxiv.org/content/early/2025/06/24/2024.08.29.610374}.

\bibitem[Wueseke et~al.(2016)Wueseke, Zwicker, Schwager, Wong, Oegema,
  J{\"u}licher, Hyman, and Woodruff]{Wueseke2016}
Oliver Wueseke, David Zwicker, Anne Schwager, Yao~Liang Wong, Karen Oegema,
  Frank J{\"u}licher, Anthony~A. Hyman, and Jeffrey~B. Woodruff.
\newblock Polo-like kinase phosphorylation determines caenorhabditis elegans
  centrosome size and density by biasing spd-5 toward an assembly-competent
  conformation.
\newblock \emph{Biology Open}, 5:\penalty0 1431--1440, 10 2016.
\newblock ISSN 20466390.
\newblock \doi{10.1242/bio.020990}.

\bibitem[Nakajo et~al.(2022)Nakajo, Kano, Tsuyama, Haruta, and
  Sugimoto]{Nakajo2022}
Momoe Nakajo, Hikaru Kano, Kenji Tsuyama, Nami Haruta, and Asako Sugimoto.
\newblock Centrosome maturation requires phosphorylation-mediated sequential
  domain interactions of spd-5.
\newblock \emph{Journal of Cell Science}, 135\penalty0 (8):\penalty0 259025, 04
  2022.
\newblock ISSN 0021-9533.
\newblock \doi{10.1242/jcs.259025}.
\newblock URL \url{https://doi.org/10.1242/jcs.259025}.

\bibitem[Rios et~al.(2024)Rios, Bagnucka, Ryder, Ferreira~Gomes, Familiari,
  Yaguchi, Amato, Stachera, Joachimiak, and Woodruff]{Rios2024}
Manolo~U. Rios, Ma{\l}gorzata~A. Bagnucka, Bryan~D. Ryder, Beatriz
  Ferreira~Gomes, Nicole~E. Familiari, Kan Yaguchi, Matthew Amato, Weronika~E.
  Stachera, {\L}ukasz~A. Joachimiak, and Jeffrey~B. Woodruff.
\newblock Multivalent coiled-coil interactions enable full-scale centrosome
  assembly and strength.
\newblock \emph{Journal of Cell Biology}, 223\penalty0 (4):\penalty0
  e202306142, 03 2024.
\newblock ISSN 0021-9525.
\newblock \doi{10.1083/jcb.202306142}.
\newblock URL \url{https://doi.org/10.1083/jcb.202306142}.

\bibitem[Young et~al.(2025)Young, Herrera, Zhang, Farhadifar, and
  Shelley]{Young2025}
Yuan-Nan Young, Vicente~Gomez Herrera, Huan Zhang, Reza Farhadifar, and
  Michael~J. Shelley.
\newblock Geometric model for dynamics of motor-driven centrosomal asters.
\newblock \emph{Phys. Rev. Res.}, 7:\penalty0 013004, Jan 2025.

\bibitem[Garcia-Baucells et~al.(2025)Garcia-Baucells, Bevilacqua, Rufin,
  Rumpf-Kienzl, Zampetaki, Andriotis, Thurner, Prevedel, F{\"u}rthauer, and
  Dammermann]{GarciaBaucells2025}
J{\'u}lia Garcia-Baucells, Carlo Bevilacqua, Manuel Rufin, Cornelia
  Rumpf-Kienzl, Alexandra Zampetaki, Orestis~G. Andriotis, Philipp~J. Thurner,
  Robert Prevedel, Sebastian F{\"u}rthauer, and Alexander Dammermann.
\newblock Centrosome softening as a mechanical adaptation for mitosis.
\newblock \emph{bioRxiv}, 2025.
\newblock \doi{10.1101/2025.09.09.675178}.
\newblock URL
  \url{https://www.biorxiv.org/content/early/2025/09/10/2025.09.09.675178}.

\bibitem[Garzon-Coral et~al.(2016)Garzon-Coral, Fantana, and
  Howard]{GarzonCoral2016}
Carlos Garzon-Coral, Horatiu~A. Fantana, and Jonathon Howard.
\newblock A force-generating machinery maintains the spindle at the cell center
  during mitosis.
\newblock \emph{Science}, 352\penalty0 (6289):\penalty0 1124--1127, 2016.
\newblock \doi{10.1126/science.aad9745}.

\bibitem[Mittasch et~al.(2020)Mittasch, Tran, Rios, Fritsch, Enos,
  Ferreira~Gomes, Bond, Kreysing, and Woodruff]{Mittasch2020}
Matth{\"a}us Mittasch, Vanna~M. Tran, Manolo~U. Rios, Anatol~W. Fritsch,
  Stephen~J. Enos, Beatriz Ferreira~Gomes, Alec Bond, Moritz Kreysing, and
  Jeffrey~B. Woodruff.
\newblock Regulated changes in material properties underlie centrosome
  disassembly during mitotic exit.
\newblock \emph{Journal of Cell Biology}, 219\penalty0 (4), 02 2020.
\newblock ISSN 0021-9525.
\newblock \doi{10.1083/jcb.201912036}.
\newblock URL \url{https://doi.org/10.1083/jcb.201912036}.

\bibitem[Banerjee and Banerjee(2025)]{Banerjee2025}
Deb~Sankar Banerjee and Shiladitya Banerjee.
\newblock Catalytic growth in a shared enzyme pool ensures robust control of
  centrosome size.
\newblock \emph{eLife}, 12:\penalty0 RP92203, feb 2025.
\newblock ISSN 2050-084X.
\newblock \doi{10.7554/eLife.92203}.

\bibitem[Cabral et~al.(2019)Cabral, Laos, Dumont, and Dammermann]{Cabral2019}
Gabriela Cabral, Triin Laos, Julien Dumont, and Alexander Dammermann.
\newblock Differential requirements for centrioles in mitotic centrosome growth
  and maintenance.
\newblock \emph{Developmental Cell}, 50\penalty0 (3):\penalty0 355--366.e6,
  2019.
\newblock ISSN 1534-5807.
\newblock \doi{10.1016/j.devcel.2019.06.004}.

\bibitem[Laos et~al.(2015)Laos, Cabral, and Dammermann]{Laos2015}
Triin Laos, Gabriela Cabral, and Alexander Dammermann.
\newblock Isotropic incorporation of spd-5 underlies centrosome assembly in c.
  elegans.
\newblock \emph{Current Biology}, 25\penalty0 (15):\penalty0 R648--R649, 2015.
\newblock ISSN 0960-9822.
\newblock \doi{10.1016/j.cub.2015.05.060}.

\bibitem[Jawerth et~al.(2020)Jawerth, Fischer-Friedrich, Saha, Wang, Franzmann,
  Zhang, Sachweh, Ruer, Ijavi, Saha, Mahamid, Hyman, and
  J{{\"u}}licher]{Jawerth2020}
Louise Jawerth, Elisabeth Fischer-Friedrich, Suropriya Saha, Jie Wang, Titus
  Franzmann, Xiaojie Zhang, Jenny Sachweh, Martine Ruer, Mahdiye Ijavi,
  Shambaditya Saha, Julia Mahamid, Anthony~A. Hyman, and Frank J{{\"u}}licher.
\newblock Protein condensates as aging {Maxwell} fluids.
\newblock \emph{Science}, 370\penalty0 (6522):\penalty0 1317--1323, 2020.

\bibitem[Takaki et~al.(2023)Takaki, Jawerth, Popovi{\'c}, and
  J{{\"u}}licher]{Takaki2023}
Ryota Takaki, Louise Jawerth, Marko Popovi{\'c}, and Frank J{{\"u}}licher.
\newblock Theory of rheology and aging of protein condensates.
\newblock \emph{PRX Life}, 1:\penalty0 013006, Aug 2023.

\bibitem[Harrington et~al.(2024)Harrington, Mezzenga, and
  Miserez]{Harrington2024}
Matthew~J Harrington, Raffaele Mezzenga, and Ali Miserez.
\newblock Fluid protein condensates for bio-inspired applications.
\newblock \emph{Nature Reviews Bioengineering}, 2\penalty0 (3):\penalty0
  260--278, 2024.

\bibitem[Sander(2020)]{sander2020}
Oliver Sander.
\newblock \emph{DUNE -- The Distributed and Unified Numerics Environment}.
\newblock Springer International Publishing, 2020.
\newblock \doi{10.1007/978-3-030-59702-3}.

\bibitem[Praetorius()]{praetorius}
Simon Praetorius.
\newblock Adaptive multi-dimensional simulations.
\newblock \url{https://gitlab.com/amdis/amdis}.

\bibitem[Witkowski et~al.(2015)Witkowski, Ling, Praetorius, and
  Voigt]{Witkowski_2015}
T.~Witkowski, S.~Ling, S.~Praetorius, and A.~Voigt.
\newblock Software concepts and numerical algorithms for a scalable adaptive
  parallel finite element method.
\newblock \emph{Advances in Computational Mathematics}, 41\penalty0
  (6):\penalty0 1145--1177, 2015.
\newblock \doi{10.1007/s10444-015-9405-4}.
\newblock URL \url{https://doi.org/10.1007/s10444-015-9405-4}.

\bibitem[Carvalho et~al.(2011)Carvalho, Olson, Gutierrez, Zhang, Noble, Zanin,
  Desai, Groisman, and Oegema]{Carvalho2011}
Ana Carvalho, Sara~K. Olson, Edgar Gutierrez, Kelly Zhang, Lisa~B. Noble,
  Esther Zanin, Arshad Desai, Alex Groisman, and Karen Oegema.
\newblock Acute drug treatment in the early c. elegans embryo.
\newblock \emph{PLOS ONE}, 6\penalty0 (9):\penalty0 1--8, 09 2011.
\newblock \doi{10.1371/journal.pone.0024656}.
\newblock URL \url{https://doi.org/10.1371/journal.pone.0024656}.

\bibitem[Essex et~al.(2009)Essex, Dammermann, Lewellyn, Oegema, and
  Desai]{Essex2009}
Anthony Essex, Alexander Dammermann, Lindsay Lewellyn, Karen Oegema, and Arshad
  Desai.
\newblock Systematic analysis in caenorhabditis elegans reveals that the
  spindle checkpoint is composed of two largely independent branches.
\newblock \emph{Molecular biology of the cell}, 20\penalty0 (4):\penalty0
  1252--1267, 2009.
\newblock \doi{https://doi.org/10.1091/mbc.e08-10-1047}.

\bibitem[Green et~al.(2011)Green, Kao, Audhya, Arur, Mayers, Fridolfsson,
  Schulman, Schloissnig, Niessen, Laband, et~al.]{Green2011}
Rebecca~A Green, Huey-Ling Kao, Anjon Audhya, Swathi Arur, Jonathan~R Mayers,
  Heidi~N Fridolfsson, Monty Schulman, Siegfried Schloissnig, Sherry Niessen,
  Kimberley Laband, et~al.
\newblock A high-resolution c. elegans essential gene network based on
  phenotypic profiling of a complex tissue.
\newblock \emph{Cell}, 145\penalty0 (3):\penalty0 470--482, 2011.
\newblock \doi{10.1016/j.cell.2011.03.037}.

\bibitem[Maddox et~al.(2006)Maddox, Portier, Desai, and Oegema]{Maddox2006}
Paul~S. Maddox, Nathan Portier, Arshad Desai, and Karen Oegema.
\newblock Molecular analysis of mitotic chromosome condensation using a
  quantitative time-resolved fluorescence microscopy assay.
\newblock \emph{Proceedings of the National Academy of Sciences}, 103\penalty0
  (41):\penalty0 15097--15102, 2006.
\newblock \doi{10.1073/pnas.0606993103}.

\end{thebibliography}

\balancecolsandclearpage

\onecolumngrid

\appendix

\section{Phase-field simulations}

\subsection{Phase-field formulation of condensate growth} \label{sec:phase_field_formulation}

In this appendix, we introduce a phase-field formulation of the sharp interface model developed in the main text, describing growth of a viscoelastic condensate. 
We treat the scaffold volume fraction as a phase-field order parameter, such that $\phi_S\sim0$ outside of the condensate, and augment the continuity equation to extend over the whole spatial domain of the cell. 
We then use the Gibbs-Duhem relation $\nabla \Pi = \phi_S\nabla\mu_S$ to rewrite the osmotic pressure in terms of chemical potential of the scaffold $\mu_S$.
This formalism approximates the sharp interface between the condensate and the surrounding environment with a diffuse interface over a finite spatial width.
We capture the evolution of this interface by introducing a free energy functional of the form
\begin{align}
    F = \int  \left[ f(\phi_P, \phi_S) + \frac{\kappa}{2}(\nabla \phi_P)^2 + \frac{\kappa}{2}(\nabla \phi_S)^2 \right] \textrm{d}V \; , \label{eq:energy}
\end{align}
where
\begin{equation} \label{app:FEDensity}
    f(\phi_P, \phi_S) = \frac{a}{2}\phi_P^2+\frac{b}{2(\phi_S^0)^2}(\phi_S-\phi_S^0)^2(\phi_S)^2 \;
\end{equation}
corresponds to a double well potential with minima centered on $\phi_S=0$ and $\phi_S=\phi_S^0$. Moreover, $\kappa={3\gamma\epsilon}/{(\phi_S^0)^2}$ and $a=b={12\gamma}/{(\phi_S^0)^2\epsilon}$, where $\epsilon$ and $\gamma$ denote the width and the surface tension of the diffuse interface, respectively. The chemical potential of the scaffold is then calculated via the functional derivative $\mu_S = {\delta F}/{\delta \phi_S}$.
In general, flux of precursor material may be driven by gradients in precursor chemical potential. For the free energy density prescribed in \Eqref{app:FEDensity}, this flux corresponds to simple diffusion.

The interface between the condensate and the environment can be associated with an intermediate level set (${\phi_S = \phi_S^0/2}$) of the scaffold density function. From the scaffold density we compute the characteristic function
\begin{equation}
    \tilde{\phi}(t, x) = \frac{1}{2} \left(1 - \tanh \left(\frac{\text{dist}(x, \phi_S(t))}{\epsilon}\right)\right)
\end{equation}
at time $t$, where $\text{dist}(x, \phi_S(t))$ denotes the signed distance of a point $x$ to the interface.
The reactions in the two bulk phases are then given by
\begin{equation}
    s = k_+ \tilde{\phi} \phi_P - k_- \phi_S \; .
\end{equation}
Following \citet{mokbel2018phase} condesate mechanics are coupled to the phase field by defining the elastic moduli as phase-dependent quantities,
\begin{align*}
    \KB(\tilde{\phi}) &= \KB \tilde{\phi}\;, \\
    \KS(\tilde{\phi}) &= \KS \tilde{\phi}\;.
\end{align*}
As a result, the elastic stress vanishes in the surrounding phase $(\tilde{\phi}=0$) and follows the previously defined Neo-Hookean law in the condensate phase $(\tilde{\phi}=1)$,
\begin{equation}
    \mathbf{\sigma} = \KB\tilde{\phi} (J_e - 1)\mathbf{I} + \KS\tilde{\phi} J_e^{-\frac{5}{3}}\left(\mathbf{B}-\frac{1}{3}\text{tr}(\mathbf{B})\mathbf{I}\right)\;.
\end{equation}
Similarly, strain relaxation occurs on different time scales: $\tau_+$ inside the condensate and $\tau_-$ in the surrounding environment. This formulations leads to the diffuse form of the strain evolution equation
\begin{equation}
    \overset{\kern-0.1em\smalltriangledown}{\mathbf{B}} = - \frac{\tilde{\phi}}{\tau_+} \left( \mathbf{B}  - \mathbf{I}\right) - \frac{1 - \tilde{\phi}}{\tau_-} \left( \mathbf{B}  - \mathbf{I}\right) \;, 
\end{equation}
where the dependence of the relaxation time on phase identity introduces phase dependence into the strain dynamics. In the sharp-interface limit, elastic stresses do not exist in the surrounding environment, which we account for by assuming rapid stress relaxation in the here ($\tau_-$ very small value). Consequently, scaffold motion is effectively restricted to the condensate region, with the scaffold velocity given by
\begin{equation}
     \vect{v} = \left( -m_S \nabla \mu_S + \frac{Em_S}{\phi_S} \nabla \cdot \boldsymbol{\sigma} \right) \tilde{\phi}\;, \label{eq:strain}
\end{equation}
where $E$ is a constant, dimensionless factor that scales the elastic force response relative to velocity, and $m_S$ is a constant scaffold mobility coefficient, such that $m_S \phi_S$ represents the effective scaffold mobility. Analogously, $m_P$ denotes the constant mobility coefficient of the precursor, and $m_P \phi_P$ its effective mobility.

\subsection{Summary of spherically symmetric equations}

We now introduce a radially symmetric representation of the phase-field formulation derived above. We express the system of equations \eqref{eq:energy} -- \eqref{eq:strain} in spherical coordinates $(r,\theta, \varphi)$, where $r$ represents the radial distance, and $\theta$ and $\varphi$ are the angular coordinates. In the radially symmetric case, the solutions depend only on the radial coordinate $r$ and time $t$, reducing the complexity of the system while preserving the essential physical characteristics and dynamics of growth.
Such a symmetry reduces the number of independent stress and strain components to two: the radial and tangential normal components. %
At the surface of the active core ($r=\rc$), we impose the following boundary conditions:
\begin{alignat}{2}
    \begin{rcases}  
    -\partial_r \mu_P &= - \frac{q}{m_P}\\
    -\partial_r \mu_S &= \frac{q \phi_P}{m_S \phi_S} \\
    v &= q \phi_P/\phi_S  \\
    \partial_r B_{rr} &= 0\\
    \partial_r B_{\theta\theta} &= 0 
    \end{rcases} & \text{ at } r = \rc\;.
\end{alignat}
Recall that $q$ denotes the rate of core-localised conversion of precursor into scaffold material. Neumann boundary conditions are applied to the precursor and scaffold densities to account for their dynamic exchange, while a Dirichlet condition on the scaffold velocity $v$ ensures mass conservation.
Finally, the characteristic function is now given by
\begin{equation}
    \tilde{\phi}(t, r) = \frac{1}{2} \left(1 - \tanh \left(\frac{r-r_\text{int}(\phi_S(t))}{\epsilon}\right)\right)\;,
\end{equation}
where $r_\text{int}(\phi_S(t))$ denotes the coordinate of the interface defined by $\{\phi_S=\phi_S^0/2\}$.
In summary, the governing equations for spherically symmetric condensate growth are then given by
\begin{align}
    r^2 \partial_t \phi_P  &= m_P \partial_r\left(r^2 \phi_P \partial_r \mu_P \right) - r^2 s\;, \label{app:SphericalStart} \\
    r^2 \partial_t \phi_S  &= m_S \partial_r \left( r^2 \phi_S \partial_{r} \mu_S \right) - E m_S \partial_r  \left( r^2 \left(\partial_r \sigma_{rr} + \frac{2}{r} (\sigma_{rr} - \sigma_{\theta\theta}) \right) \right) + r^2 s\;, \\
    \mu_P  &= a \phi_P  - \kappa \left(\partial_{rr} \phi_P + \frac{2}{r} \partial_r \phi_P \right)\;, \\
    \mu_S  &= \frac{b}{(\phi_s^0)^2} \left( 2 \phi_S^3 - 3 \phi_S^0 \phi_S^2 + (\phi_s^0)^2 \phi_S \right) - \kappa \left(\partial_{rr} \phi_S + \frac{2}{r} \partial_r \phi_S \right)\;, \\
    s  &= k_{+} \tilde{\phi} \phi_P  - k_{-} \phi_S\;, \\
    v  &= -\tilde{\phi}m_S \partial_r \mu_S  + \tilde{\phi} \frac{E m_S}{\phi_S} \left(\partial_r \sigma_{rr} + \frac{2}{r} (\sigma_{rr} - \sigma_{\theta\theta}) \right)\;, \\
    \partial_t B_{rr} &= - v \partial_r B_{rr} + 2 B_{rr} \partial_r v  -\frac{\tilde{\phi}}{\tau_+} \left( B_{rr} - 1 \right)  - \frac{1 - \tilde{\phi}}{\tau_-} \left( B_{rr} - 1 \right)\;, \\
    \partial_t B_{\theta\theta} &= - v \partial_r B_{\theta\theta} + 2 B_{\theta\theta} \frac{1}{r} v -\frac{\tilde{\phi}}{\tau_+} \left( B_{\theta\theta} - 1 \right)  - \frac{1 - \tilde{\phi}}{\tau_-} \left( B_{\theta\theta} - 1 \right)\;, \\
    \sigma_{rr}  &= \KB \tilde{\phi} (J-1)  + \KS \tilde{\phi} J^{-\frac{5}{3}}(B_{rr} - \frac{1}{3} (B_{rr}+2B_{\theta\theta}))\;, \\
    \sigma_{\theta\theta}  &= \KB \tilde{\phi} (J-1)  + \KS \tilde{\phi} J^{-\frac{5}{3}}(B_{\theta\theta} - \frac{1}{3} (B_{rr}+2B_{\theta\theta}))\;, \\
    J &= \sqrt{B_{rr}B_{\theta\theta}^2}\;. \label{app:SphericalEnd}
\end{align}

\subsection{Details of numerical method}

To numerically solve the above equations, we use the finite element method, utilising the DUNE~\citep{sander2020} and AMDiS~\citep{praetorius, Witkowski_2015} toolboxes.
We adopt a semi-implicit time discretisation to solve the coupled system at the $n$-th time step. 
We divide the time interval $\left[0, T\right]$ into $N$ subintervals of size $\Delta t$, such that the $n$-th time step is given by $t^n := n \cdot \Delta t$. For the spatial discretisation, we use Lagrange P2 elements for the precursor and scaffold densities, $\phi_P$ and $\phi_S$, as well as for the chemical potentials, $\mu_P$ and $\mu_S$. Linear Lagrange (P1) elements are used to approximate the scaffold velocity $v$, the stresses $\sigma_{rr}$ and $\sigma_{\theta\theta}$, and the strains $B_{rr}$ and $B_{\theta\theta}$. 
We now present the complete discrete system of equations in the weak form. The finite element spaces
\begin{align*}
    {C_h}_1 &= \left\{ c \in C^0(\bar{\Omega}) \mid c|_k \in P_2(k), k \in T_\Omega \right\}\;, \\
    {C_h}_2 &= \left\{ c \in C^0(\bar{\Omega}) \mid c|_k \in P_1(k), k \in T_\Omega \right\}\;, \\
    C_{q,n} &= \left\{ c \in C^0(\bar{\Omega}) \mid c|_k \in P_1(k), k \in T_\Omega, c(\rc)=q\phi_P^{n-1}(\rc)/\phi_S^{n-1}(\rc), c(r\sub{d})=0 \right\}
\end{align*}
are used, where $T_\Omega$ is the triangulation of the one dimensional domain $\Omega = [\rc, r\sub{d}]$.
To enhance readability, we present the equations individually, even though they are part of a unified coupled system.

The precursor density equation is discretised semi-implicitly in time taking $\phi_P^{n-1}$ and $\phi_S^{n-1}$ from the previous time step to calculate the mobility term and the characteristic function $\tilde{\phi}^{n-1}$. This leads to the weak form: find $(\phi_P^n, \mu_P^n) \in {C_h}_1 \times {C_h}_1$ such that for all $(w_1, w_2) \in {C_h}_1 \times {C_h}_1$ holds
\begin{align*}
    \int_{\Omega} r^2\frac{\phi_P^n - \phi_P^{n-1}}{\Delta t}w_1 \diff r &= 
    - m_P \int_\Omega r^2 \phi_P^{n-1} \partial_r \mu_P^n \partial_r w_1 \diff r
    + q r^2 \phi_P^{n-1}(r_c) w_1(r_c) \\
    &\hspace{1em}- k_+\int_\Omega r^2 \tilde{\phi}^{n-1} \phi_P^{n} w_1 \diff r + k_-\int_\Omega r^2 \phi_S^{n} w_1 \diff r,\; \\
    \int_\Omega \mu_P^n w_2 \diff r &= \int_\Omega a \phi_P^n w_2 \diff r + \kappa \int_\Omega \partial_r \phi_P^n \partial_r w_2 \diff r - \kappa \int_\Omega \frac{2}{r} \partial_r \phi_P^n w_2 \diff r\;. %
\end{align*}
The semi-implicit time discretisation of the scaffold density equation uses $\phi^{n-1}$ and $\tilde{\phi}^{n-1}$ from the previous time step, but the stress from the current time step. The weak form reads: find $(\phi_S^n, \mu_S^n) \in {C_h}_1 \times {C_h}_1$ such that for all $(w_3, w_4) \in {C_h}_1 \times {C_h}_1$ holds
\begin{align*}
    \int_{\Omega} r^2 \frac{\phi_S^n - \phi_S^{n-1}}{\Delta t}w_3 \diff r 
    &= -m_S \int_\Omega r^2 \phi_S^{n-1} \partial_r \mu_S^n \partial_r w_3 \diff r - q r^2 \phi_P^{n-1}(r_c) w_1(r_c) \\
    &\hspace{1em}+ E m_S \int_\Omega  r^2 \left(\partial_r \sigma_{rr}^n + \frac{2}{r} (\sigma_{rr}^n - \sigma_{\theta\theta}^n) \right) \partial_r w_3 \diff r + k_+\int_\Omega r^2 \tilde{\phi}^{n-1} \phi_P^{n} w_3 \diff r - k_-\int_\Omega r^2 \phi_S^{n} w_3 \diff r\;, \\
    \int_\Omega \mu_S^n w_4 \diff r 
    &= \int_\Omega \frac{b}{(\phi_s^0)^2} \left( 2 (\phi_S^n)^3 - 3 \phi_S^0   (\phi_S^n)^2 \right) w_4 \diff r + \int_\Omega b   \phi_S^n w_4 \diff r \\
    &\hspace{1em}+ \kappa \int_\Omega \partial_r \phi_S^n \partial_r w_4 \diff r - \kappa \int_\Omega \frac{2}{r} \partial_r \phi_S^n w_4 \diff r\;. %
\end{align*}
The nonlinear terms in the last equation are linearised using a first-order Taylor approximation, but we refrain from explicitly writing out the linearised terms for the sake of readability. The evolution of scaffold material is directly coupled to that of precursor material through the reaction term $s$, and to the mechanics via the stress. 
We use the semi-implicit time discrete form of the strain equations taking $\tilde{\phi}^{n-1}$ from the previous time step, but the scaffold velocity and the strains from the current time step. The weak form reads: find $(B_{rr}^n, B_{\theta\theta}^n) \in {C_h}_2 \times {C_h}_2$ such that for all $(w_5, w_6) \in {C_h}_2 \times {C_h}_2$ holds
\begin{align*}
    \int_\Omega \frac{B_{rr}^n - B_{rr}^{n-1}}{\Delta t}w_5 \diff r &= \int_\Omega \left(- v^n\partial_rB_{rr}^{n} +  2B_{rr}^{n}\partial_r v^n \right) w_5 \diff r  - \int_\Omega \frac{\tilde{\phi}^{n-1}}{\tau_+} \left( B_{rr}^n - 1 \right)w_5 \diff r  - \int_\Omega \frac{1 - \tilde{\phi}^{n-1}}{\tau_-} \left( B_{rr}^n - 1 \right)w_5 \diff r\;, \\
    \int_\Omega \frac{B_{\theta\theta}^n - B_{\theta \theta}^{n-1}}{\Delta t}w_6 \diff r &= \int_\Omega \left(- v^n\partial_rB_{\theta\theta}^{n} + \frac{2}{r}B_{\theta \theta}^{n} v^n \right) w_6 \diff r - \int_\Omega \frac{\tilde{\phi}^{n-1}}{\tau_+} \left( B_{\theta\theta}^n - 1 \right)w_6 \diff r  - \int_\Omega \frac{1 - \tilde{\phi}^{n-1}}{\tau_-} \left( B_{\theta\theta}^n - 1 \right)w_6 \diff r\;.
\end{align*}
The nonlinear terms in the strain equations are linearised using a first-order Taylor approximation.
The velocity equation is discretised semi-implicitly in time using the characteristic function $\tilde{\phi}^{n-1}$ from the previous time step. Similarly, the stress equations are discretised semi-implicitly, incorporating $\tilde{\phi}^{n-1}$ from the previous time step along with the current stresses and strains. The weak form reads: find $(v_r^n, \sigma_{rr}^n, \sigma_{\theta\theta}^n) \in C_{q,n} \times {C_h}_2 \times {C_h}_2$ such that for all $(w_7, w_8, w_9) \in  C_{0,n} \times {C_h}_2 \times {C_h}_2$ holds
\begin{align*}
    \int_\Omega &v^n w_7 \diff r = -m_S \int_\Omega \tilde{\phi}^{n-1} \partial_r \mu_S^n w_7 \diff r + E m_S \int_\Omega \frac{\tilde{\phi}^{n-1}}{\phi_S^{n-1}} \left(\partial_r \sigma_{rr}^n + \frac{2}{r} \sigma_{rr}^n - \frac{2}{r} \sigma_{\theta\theta}^n\right)w_7 \diff r\;, \\
    \int_\Omega &\sigma_{rr}^n w_8 \diff r = \int_\Omega \KB \tilde{\phi}^{n-1} \left(\left(B_{rr}^n\right)^{1/2} B_{\theta\theta}^n -1\right)w_8 \diff r + \int_\Omega \KS \tilde{\phi}^{n-1} \left(B_{rr}^{n}\left(B_{\theta\theta}^{n}\right)^{2}\right)^{-\frac{5}{6}}\frac{2}{3}\left(B_{rr}^{n} - B_{\theta\theta}^{n}\right) w_8 \diff r, \\
    \int_\Omega &\sigma_{\theta\theta}^n w_9 \diff r = \int_\Omega \KB \tilde{\phi}^{n-1} \left(\left(B_{rr}^n\right)^{1/2} B_{\theta\theta}^n -1\right)w_9 \diff r + \int_\Omega \KS \tilde{\phi}^{n-1} \left(B_{rr}^{n}\left(B_{\theta\theta}^{n}\right)^{2}\right)^{-\frac{5}{6}} \frac{1}{3}\left(-B_{rr}^{n} + B_{\theta\theta}^{n}\right) w_9 \diff r\;.
\end{align*}
Again, note that the stress is coupled to the strain by $B_{rr}^n$ and $B_{\theta\theta}^n$ and to the phase-field equation by $\tilde{\phi}^{n-1}$. Nonlinear terms appear in both the isotropic and deviatoric parts of the stress equations. These are replaced by first-order Taylor expansions, which are not written out here.

In order to accurately resolve the forces at the interface we utilise an adaptive grid. The interfacial grid is refined heuristically, determined by the value of the phase field. At each time step, the position of the interface is computed by finding where $\phi_S = \phi_S^0/2$. Within an interval of fixed width around the interface, the grid is refined according to the smaller grid width $h_\text{interface}$. In a larger interval around the interface as well as inside the condensate an intermediate grid width $h_\text{transition}$ is applied. Elsewhere, the grid width is set to $h_\text{bulk}$, with $h_\text{interface} \le h_\text{transition} \le h_\text{bulk}$.
Initially, we prescribe the phase field as 
\begin{equation*}
    \phi_S(r) = \frac{\phi_S^0 }{2} \left(1 - \tanh \left( \frac{r - R_0}{\epsilon}  \right)\right) + \psi\;,
\end{equation*}
where $R_0$ denotes the radius of the initial droplet at $t=0$, and $\psi$ is an initial offset value. The value of $\psi$ is derived by analysing the total free energy of the system and is chosen such that, in the absence of reactions, neither shrinking nor growing of the droplet would lead to a decrease in energy. Specifically, we set
\begin{equation*}
    \psi = \frac{2\gamma}{R_0 b\phi_S^0}\;.
\end{equation*} 
The simulation parameters used for the phase-field results presented in Fig. 2 are provided in \tabref{tab:phase-field-parameters}.

\noindent\begin{table}
    \centering
    \caption{Overview of the simulation parameters used in the phase-field simulations.}
    \label{tab:phase-field-parameters}
    \begin{tabularx}{1\textwidth} { L{1.9} C{0.5} C{0.6}}
        Quantity & Symbol & Value \\
        \hline
        Initial radius of the condensate & $R_0$ & $\SI{1.5}{\micro\m}$ \\
        Radius of the active core  & $\rc$  & $\SI{0.3}{\micro\m}$ \\
        Right domain boundary  & $r\sub{d}$  & $\SI{20}{\micro\m}$ \\
        Domain & $\Omega$ & $[\rc, r\sub{d}]$\\
        Width of the diffuse interface & $\epsilon$ & $\SI{0.01}{\micro\m}$\\
        Grid width at the interface & $h_\text{interface}$ & $1.97\times 2^{-10} \si{\micro\m}$\\
        Intermediate grid width and grid width inside the droplet & $h_\text{transition}$ & $1.97\times 2^{-9} \si{\micro\m}$\\
        Grid width in the surrounding environment & $h_\text{bulk}$ & $1.97\cdot 2^{-3} \si{\micro\m}$\\
        Time step size & $\Delta t$ & $\SI{0.0025}{\s}$\\
        Reference volume fraction of the scaffold & $\phi_S^0$ & $\num{0.1}$\\
        Initial volume fraction of the precursor & $\phi_P^0$ & $\num{0.01}$\\
        Surface tension & $\gamma$ & $\SI{1}{\pico\N\per\micro\m}$\\
        Mobility constant of the scaffold & $m_S$ & $\SI{1e8}{\micro\m\tothe{4}\per\N\per\s}$\\
        Mobility constant of the precursor & $m_P$ & $\SI{1e12}{\micro\m\tothe{4}\per\N\per\s}$\\
        Rate of conversion of scaffold into precursor  & $k_-$ & $\SI{0.001}{\per\s}$\\
        Rate of conversion of precursor into scaffold in the bulk of the condensate & $k_+$ & $\SI{0.4}{\per\s}$\\
        Rate of conversion of precursor into scaffold at the surface of the active core& $q$ & $\SI{4.2e-8}{\per\s}$\\
        Constant scaling factor of the elastic force response relative to velocity & $E$ & $100$ \\
        Bulk modulus of the scaffold & $\KB$ & $\SI{1}{\N\per\square\m}$ \\
        Shear modulus of the scaffold & $\KS$ & $\SI{1}{\N\per\square\m}$\\
        Relaxation time in the bulk of the condensate & $\tau_-$ & $\SI{100}{\s}$\\
        Relaxation time in the surrounding environment & $\tau_+$ & $\SI{0.1}{\s}$\\
    \hline
    \end{tabularx}
\end{table}

\section{Reduced model}

\subsection{Detailed derivation of integrated model}

Integrating the continuity equation over the volume of the condensate from $\rc$ up to some radius $r$, we find
\begin{equation}\label{App:IntMassBalance}
 \pd{}{t}\int_{\rc}^r  \phi_{S} r^2\textrm{d}r+j_rr^2 =\int_{\rc}^r sr^2\textrm{d}r,   
\end{equation}
where we have used the divergence theorem to convert the flux term to a surface integral, which we then evaluate. Note that integration over the angular directions simply results in a factor of $4\pi$ multiplying each term.
We define the mean scaffold fraction and mean reaction flux as ${\bar{\phi}_s=3(R^{3}-\rc^3)^{-1}\int_{\rc}^{R}\phi_Sr^2\textrm{d}r}$ and ${\bar s=3(R^{3}-\rc^3)^{-1}\int_{\rc}^{R}\left(k_{+}\bar{\phi}_{P}-k_{-} {\phi}_{S} \right)r^2\textrm{d}r}$, respectively.
Assuming that $\phi_{S}$ varies only weakly with $r$, such that $ \phi_{S}(r)=\bar{\phi}_S+\epsilon(r)$, where $\epsilon\ll\bar{\phi}_S$, we then write \Eqref{App:IntMassBalance} as
\begin{equation}
    \frac{r^3-\rc^3}{3}\dd{\bar{\phi}_S}{t}+j_rr^2=\frac{r^3-\rc^3}{3}\bar{s}+q\bar{\phi}_{P} \rc^2.
\end{equation}
Rearranging, we then arrive at an explicit expression for the flux profile within the condensate,
\begin{equation}\label{App:jExpression}
    j_r(r)=\frac{r^3-\rc^3}{3r^2}\left[\bar s-\dd{\bar{\phi}_S}{t}\right]+\frac{q\bar{\phi}_{{P}}\rc^2}{r^2} \; .
\end{equation}
Applying the kinematic boundary condition $\dd{R}{t}=v(R)={j(R)}/{ \bar{\phi}_{S}}$ at the centrosome surface then gives
\begin{equation}\label{App:eq:dRdt}
    \dd{R}{t}=\frac{R^3-\rc^3}{3 \bar{\phi}_{S}R^2}\left[\bar{s}-\dd{ \bar{\phi}_{S}}{t}\right]+\frac{q\bar{\phi}_{P}\rc^2}{ \bar{\phi}_{S} R^2},
\end{equation}
which is an evolution equation for $R$ as a function of $\bar{\phi}_{S}(t)$ and $\bar{\phi}_{P}(t)$. Note that $\bar{\phi}_{P}(t)$ can be calculated directly via conservation of total condensate material,
$ V(\bar{\phi}_{S}+\bar{\phi}_{P})+\chi\bar{\phi}_{P}(\Vc-V)=\bar{\phi}\Vc$,
where we recall that $\bar{\phi}$ is the mean volume fraction of condensate material (scaffold plus precursor) in the cell, and $\chi$ quantifies any preferential partitioning of precursor material into the condensate.

The total stress can be decomposed into different components arising from elasticity, hydrostatic pressure  and osmotic potential, which we denote by $\boldsymbol{\sigma}$, $p$ and $\Pi$, respectively. Neglecting inertia and the influence of body forces (\textit{e.g.}, gravity), conservation of momentum then requires that
\begin{equation}\label{App:TotalStressBalance}
    \nabla\cdot\mathbf{T}=0,
\end{equation}
where $\mathbf{T}$ is the \textit{total} stress given by
\begin{equation}
   \mathbf{T}=\boldsymbol{\sigma}-(p+\Pi)\mathbf{I},
\end{equation}
and $\mathbf{I}$ is the identity tensor.
To capture the effect of external forces, we assume that there exists a surface stress~$\sigma^*$, which acts in a radial direction at the boundary of the condensate. As a result, the boundary of the condensate is constrained to satisfy the kinematic condition 
\begin{equation}\label{App:StressBC}
    -\mathbf{T}\cdot\hat{\boldsymbol{r}}=\sigma^*
\end{equation}
at $r=R$.
Note that the magnitude of $\sigma^*$ may in principle depend on $R$. 
In the absence of elastic stresses, the osmotic pressure acts to set an energetically preferred value for the scaffold volume fraction, denoted by $ \phi_{S}^0$. Defining the difference in $\Pi$ between this base state and the full system (including elasticity) as $\Delta\Pi=\Pi( \phi_{S})-\Pi( \phi_{S}^0)$, we can thus rewrite \eqref{App:TotalStressBalance} in spherical coordinates as
\begin{equation}\label{App:StressBalanceRadial}
    \pd{}{r}(\sigma_{rr}-\Delta\Pi)+2\left(\frac{\sigma_{rr}-\sigma_{\theta\theta}}{r}\right)=0,
\end{equation}
where $\sigma_{rr}$ and $\sigma_{\theta\theta}$ are the radial and azimuthal normal components of $\boldsymbol{\sigma}$.
Integrating \eqref{App:StressBalanceRadial} from $r=r$ to $r=R$, and substituting the boundary condition \eqref{App:StressBC} then gives
\begin{equation}\label{App:IntegratedStressBalance}
    \Delta\Pi[ \phi_{S}(r)]-\sigma_{rr}(r)-\sigma^*+\int_r^R\frac{2}{r'}(\sigma_{rr}-\sigma_{\theta\theta})\textrm{d}r'=0.
\end{equation}
Assuming that $ \phi_{S}$ does not deviate significantly from its preferred value $ \phi_{S}^0$, we can linearise $\Delta\Pi$ such that ${\Delta\Pi\sim\alpha^{-1}( \phi_{S}- \phi_{S}^0)}$, where $\alpha$ is a small parameter that depends on the specific energetic interactions between the scaffold and its pore fluid. Rearranging \eqref{App:IntegratedStressBalance} then gives an explicit expression for $ \phi_{S}(r)$,
\begin{equation}
     \phi_{S}(r)= \phi_{S}^0+\alpha\left[\sigma_{rr}-\int_r^R\frac{2}{r'}(\sigma_{rr}-\sigma_{\theta\theta})\textrm{d}r' +\sigma^*\right].
\end{equation}
Finally, averaging $ \phi_{S}(r)$ over the volume of the centrosome, we arrive at an expression for the mean scaffold density as a function of elastic stress and microtubule pulling force,
\begin{equation}\label{App:eq:phiEquation}
     \bar{\phi}_{S}= \phi_{S}^0+\frac{3\alpha}{R^3-\rc^3}\int_{\rc}^R\textrm{tr}(\boldsymbol{\sigma})r^2\textrm{d}r+\sigma^*(R),
\end{equation}
where $\textrm{tr}(\boldsymbol{\sigma})=\sigma_{rr}+2\sigma_{\theta\theta}$ can be calculated directly from the elastic stress tensor.

The reduced oder model derived here comprises a non-linear, coupled ODE--PDE--algebraic system. We solve this system of equations numerically in MATLAB by rescaling to a fixed domain $r_*=(r-\rc)/(R-\rc)$. We approximate spatial derivatives by a finite volume scheme, and use MATLAB's in-built solver \texttt{ode15i} for timestepping.

\subsection{Steady-state solutions}

For an uncoupled system, in which elastic stresses do not affect the scaffold density, we can straightforwardly find analytical expressions for the steady state condensate volume $\Vf$ by setting the right side of \Eqref{App:eq:dRdt} equal to zero. As in the main text, we set $\chi=1$ for simplicity such that $\bar{\phi}_P=\bar{\phi}-\phi_S^0\Vf/\Vc$. If incorporation only occurs at the core ($k=0$) we then find
\begin{equation}
\Vf = \frac{\bar{\phi}\Vc}{\phi_S^0\left(1+ k_- \Vc/4\pi q \rc^2 \right)} \; .
\end{equation}
For a general reaction scheme including both core and bulk incorporation, we find
\begin{equation}
\Vf = \frac{\bar{\phi}\Vc}{2\phi_s^0}\left[ \left(1- \frac{k_-\phi_S^0}{k_+\bar{\phi}} - \frac{4\pi q\rc^2\phi_s^0}{k_+\bar{\phi}\Vc}\right)+\sqrt{\left(1- \frac{k_-\phi_S^0}{k_+\bar{\phi}} - \frac{4\pi q\rc^2\phi_s^0}{k_+\bar{\phi}\Vc}\right)^2+\frac{16\pi q\rc^2\phi_s^0}{k_+\bar{\phi}\Vc}} \right] \;.
\end{equation}

\subsection{Deformation of a viscoelastic sphere}

In the limits $\tau\to0$ (fast strain relaxation) and $\tau\to\infty$ (no strain relaxation), we can find approximate analytical expressions for the change in condensate volume resulting from the sudden introduction of a constant external stress $\sigma^*$. In the absence of new material incorporation (reactions complete), conservation of mass implies that
$V\bar{\phi}_S=\textrm{const}$. We denote the condensate volume before the external stress is applied as $V_0$, and assume that the condensate is unstrained ($\bar{\phi}_S={\phi}_S^0$) at this point. In the limit $\tau\to0$, internal elastic stresses dissipate instantaneously, and so $\bar{\phi}_S=\phi_S^0+\alpha\sigma^*$. The normalised change in volume $\Delta V = \left(V-V_0\right)/V_0$ resulting from the external stress $\sigma^*$ is then
\begin{equation}\label{eq:AppDefLiq}
{\Delta V} = -\frac{\alpha\sigma^*}{\phi_0+\alpha\sigma^*}\;.
\end{equation}
To analyse the limit $\tau\to\infty$, we note that if $\Vcore\ll V_0$ both the radial and azimuthal normal strains are approximately uniform, and equal to each other. We thus write $B=B_{rr}=B_{\theta\theta}$.
Correspondingly, the elastic pressure is then $\bar{\sigma}=\mathcal{K}(B^{\frac{3}{2}}-1)$ for a Neo-Hookean constitutive model.
For a uniformly stretched material, strain is related to volume by $B=(V/V_0)^{2/3}$.
As such, we can use that $\phi_S=\phi_S^0+\alpha\sigma+\alpha\sigma^*$ to find
\begin{equation}\label{eq:AppDefSol}
\Delta V = -\frac{1}{2}\left(1+\frac{\sigma^*}{{K}}+\frac{\phi_0}{\alpha{K}}\right) + \frac{1}{2}\sqrt{\left(1+\frac{\sigma^*}{{K}}+\frac{\phi_0}{\alpha{K}}\right)^2-\frac{4\sigma^*}{{K}}}\;.
\end{equation}
Finally, we can rewrite \Eqsref{eq:AppDefLiq} and \eqref{eq:AppDefSol} as expressions for the external stress required to induce a certain deformation $\Delta V$.
In the limit $\tau\to 0$, we have
\begin{equation}
\alpha\sigma^* = \frac{\phi_S^0\Delta V}{1+\Delta V}\;,
\end{equation}
and in the limit $\tau\to \infty$, we have
\begin{equation}
\alpha\sigma^* = \frac{\phi_S^0\Delta V}{1+\Delta V} + \alpha K \Delta V\; .
\end{equation}

\section{Experimental details}

The quantitative analysis of SPD-5 PCM scaffold polymer expansion in one-, two- and four-cell stage \textit{C. elegans} embryos presented in Fig. 6 is based on experimental data described in the accompanying manuscript by \citet{GarciaBaucells2025}. We briefly outline the experimental methods below. For further details on the methods used please refer to that manuscript.

\subsection{Materials and methods}

Experiments were performed in worms co-expressing GFP:SPD-5, mCherry:SPD-2 and mCherry:Histone (strain DAM946: \textit{spd-5(vie26[gfp::spd-5 +loxP]) I; unc-119(ed3) III; ltIs37[pAA64; Ppie-1::mcherry::his-58; unc-119(+)]; ltIs69[pAA191; Ppie-1::mcherry::spd-2; unc-119(+)] IV}~\citep{Cabral2019}. To enable nocodazole-treatment of early embryos, egg shells were permeabilised using \textit{perm-1} RNAi~\citep{Carvalho2011}, while simultaneously inhibiting the spindle checkpoint with \textit{${\textrm{mad-}2^{\textrm{mdf-2}}}$} RNAi to prevent mitotic arrest~\citep{Essex2009}. RNAi was performed by soaking~\citep{Green2011}, with a recovery time of 24 hours at 20 °C to minimise phenotypes beyond embryo permeability. 

\subsection{Imaging}

Live cell imaging was performed using a Yokogawa CSU-X1 spinning disk confocal mounted on a Zeiss Axio Observer Z1 microscope equipped with 100 mW 488 nm and 561 nm solid-state lasers and an Excelitas pco.edge 4.2 sCMOS camera and controlled by VisiView 6.0 software (Visitron Systems). For the imaging in Fig. 6D, young adult animals were dissected on 24 x 60 mm coverslips in an 8 $\upmu$l droplet of 30\% iodixanol (STEMCELL Technologies, Cat \# 07820) in M9 (refractive index matched to embryos, n = 1.365) containing 24.7 $\upmu$m polystyrene spacer beads (Polysciences, Cat \# 07313-5) to minimise spherical aberration during imaging. Samples were mounted under an 18 x 18 mm coverslip and sealed with two-component dental silicone (Picodent eco-sil speed). Data were acquired using a 63x 1.3 NA LCI Plan-Neofluar glycerol/water immersion objective with DIC III optics and correction collar, optimised for the refractive index of the mounting medium. 33 x 0.5 $\upmu$m GFP z-stacks along with single mid-plane mCherry and DIC images were acquired every 15s, using low laser illumination to minimise photobleaching.

\subsection{Nocodazole treatment} 

For the nocodazole treatment experiments in Fig. 6E, \textit{perm-1;mdf-2} double RNAi-treated worms were dissected onto 24 x 60 mm coverslips in an 8 $\upmu$l droplet of meiosis medium (60\% Leibovitz's L-15 medium, 25 mM HEPES pH 7.4, 0.5\% inulin, 20\% heat-inactivated feotal bovine serum) containing 15.7 $\upmu$m polystyrene spacer beads (Polysciences, Cat \# 18328-5) and covered with an 18 x 18 mm coverslip, using the slight compression to immobilise embryos between the two coverslips. The sample was sealed with silicone on two sides only to permit medium exchange and imaged as before except using a 63x 1.4 NA Plan-Apochromat oil immersion objective with DIC III optics. For drug application in S-phase (post-centrosome separation) or at NEBD, meiosis medium was replaced with medium containing 0.01 $\upmu$g/$\upmu$l nocodazole at the indicated time point by drawing liquid through the flow chamber. Embryos requiring $>$ 3 min for medium exchange were excluded from further analysis. Controls for nocodazole-treatment experiments were \textit{perm-1;mdf-2} double RNAi embryos maintained in drug-free meiosis medium. 

\subsection{Image processing}

Post image acquisition, image stacks were processed using Fiji/ImageJ (version 1.54f), separating channels and extracting timestamp metadata, with subsequent analysis performed in MATLAB (version R2021b). Movies were aligned with respect to two reference time points: (1) male pronuclear/nuclear permeabilisation for S-phase-treated centrosomes, and (2) chromosome decondensation for NEBD-treated embryos. Nuclear permeabilisation was defined as the last time point preceding complete cytoplasmic GFP invasion of the nuclear space. Chromosome decondensation timing was determined using a modified version of the method by~\citet{Maddox2006}. Prior to analysis, images were corrected for spherical aberration, camera noise and photobleaching. 

For centrosome segmentation, the centrosome mid-plane was identified as the plane with the maximum integrated intensity within a 60 pixel diameter prismatic volume centred around the centrosome. Mild Gaussian filtering (kernel size: 3 x 3, $\sigma = 1$) was then applied, followed by Otsu thresholding (using MATLAB’s \texttt{multithresh} function) for 2D segmentation of the centrosome mid-plane. Morphological properties including centroid coordinates and equivalent radius measurements were extracted using MATLAB’s \texttt{regionprops} function. For closely spaced centrosomes at the 2D mid-plane before separation, MATLAB's \texttt{watershed} function was used to resolve individual centrosomes. Centrosomes that were either truncated or out of focus were excluded based on an analysis of their 3D spatial distribution. Centrosome volume was estimated from intensity-based 2D measurements, assuming the centrosome to be a perfect sphere using $V = 4 \pi r^3/3$. The 2D radii were corrected by a fixed 0.3 $\upmu$m offset, as this was the systematic and size-independent difference identified between our 2D mid-plane and full 3D Otsu segmentation methods~\citep{GarciaBaucells2025}.

\end{document}